\documentclass{jfm}
\usepackage{epstopdf}
\usepackage{amsmath}
\usepackage{color}
\usepackage{threeparttable}

\newcommand\bu{{\boldsymbol{u}}}

\newcommand\be{{\boldsymbol{e}}}

\renewcommand\Pran{{\rm Pr}}
\newcommand\Nu{{\rm Nu}}

\begin{document}

\newtheorem{lemma}{Lemma}
\newtheorem{corollary}{Corollary}

\shorttitle{Layer Formation in Sedimentary Fingering Convection} 
\shortauthor{Reali et al} 

\title{Layer Formation in Sedimentary Fingering Convection}

\author
 {
 J. F. Reali\aff{1},
  P. Garaud\aff{1}  \corresp{\email{pgaraud@ucsc.edu}}, 
  A. Alsinan\aff{2} and E. Meiburg\aff{2}
  }

\affiliation
{
\aff{1}
Department of Applied Mathematics and Statistics, University of California, Santa Cruz, CA 95064, USA\\
\aff{2}
Department of Mechanical Engineering, University of California, Santa Barbara, CA 93106, USA\\
}

\maketitle

\begin{abstract}
When particles settle through a stable temperature or salinity gradient they can drive an instability known as sedimentary fingering convection. This phenomenon is thought to occur beneath sediment-rich river plumes in lakes and oceans, in the context of marine snow where decaying organic materials serve as the suspended particles, or in the atmosphere in the presence of aerosols or volcanic ash. Laboratory experiments of \citet{houk1973descent} and \citet{green1987importance} have shown sedimentary fingering convection to be similar to the more commonly known thermohaline fingering convection in many ways. Here, we study the phenomenon using 3D direct numerical simulations. We find evidence for layer formation in sedimentary fingering convection in regions of parameter space where it does not occur for non-sedimentary systems. This is due to two complementary effects. Sedimentation affects the turbulent fluxes and broadens the region of parameter space unstable to the $\gamma$-instability  \citep{radko2003mechanism} to include systems at larger density ratios. It also gives rise to a new layering instability that exists in $\gamma-$stable regimes. The former is likely quite ubiquitous in geophysical systems for sufficiently large settling velocities, while the latter probably grows too slowly to be relevant, at least in the context of sediments in water. 
\end{abstract}

\section{Introduction}
\label{sec:intro}

Double-diffusive convection in the so-called ``fingering" regime can occur in a fluid whose density is determined by two components which diffuse at different rates, and where the more rapidly-diffusing component is stably stratified while the more slowly-diffusing one is unstably stratified \citep{stern1960sfa,turner1974ddp,turner1985mc,schmitt1994ddo,kunze2003ros,radko2013double}. It is commonly found in a significant fraction of the upper layers of the tropical and subtropical ocean \citep{you2002goc}, where warmer and saltier water overlays cooler and fresher water. Fingering convection contributes to some degree to global diapycnal mixing \citep{Schmitt2003,schmitt2005enhanced} of the oceanic water masses. Further afield, fingering convection is also thought to occur in the interior of stars, where the fast-diffusing, stably stratified component is temperature, and where the slowly diffusing component can consist of any chemical species heavier than hydrogen \citep{vauclair2004mfa,Brownal2013,Garaud2013}. As reviewed in Section \ref{sec:intro1}, much progress has recently been made in understanding the dynamics of fingering convection, from the basic linear instability to the various phenomena associated with its nonlinear saturation (most notably, the formation of thermocompositional staircases).

Within the scope of the current investigation we focus on a closely related scenario in which the role of the unstably stratified, more slowly diffusing scalar is played by a dilute particle field, which we henceforth refer to as ``sediments." If these sediments consist of small, dense, solid, non-cohesive particles, their collective evolution can be described by an advection-diffusion equation analogous to the one governing salinity (for instance), augmented by a gravitational settling term \citep{necker2002high}. Naturally-occurring examples of this kind of system include the buoyant outflow of a warm, sediment-laden river into a colder lake, or a buoyant, sediment-laden freshwater plume into the salty ocean. Evidence from a number of different laboratory investigations \citep[e.g.][]{houk1973descent,green1987importance} suggests that the above situation gives rise to instabilities that are quite similar in nature to classical fingering instabilities. Such double-diffusive sedimentation has also been hypothesized to be active in field observations in lakes \citep{scheual15,Sanchezoget2007} and could play an important role in the rate of transport of CO$_2$ from the surface layer towards the deep ocean \citep{alldredge1987can} through the so-called ``marine snow" (i.e. slowly settling biogenic organic and inorganic material). Finally, it may also apply to hot volcanic ash plumes in the atmosphere or to the dynamics of droplets in certain classes of clouds \citep{carazzo2013particle}. 

We now begin by summarizing what is known of standard fingering convection, and then proceed to review the case of sedimentary fingering convection. 

\subsection{Standard fingering convection}
\label{sec:intro1} 

In everything that follows, we shall consider general fingering systems, but for simplicity will continue referring to the rapidly diffusing component as ``temperature" $T$, and to the slowly-diffusing component as the ``solute concentration" $C$. As such, the density of the fluid $\rho$ is assumed to be related to $T$ and $C$ through an incompressible equation of state, namely $\rho = \rho(T,C)$. As shown by \citet{baines1969}, the dynamics of fingering convection in an unbounded fluid with constant background gradients of temperature and solute concentration $T_{0z}$ and $C_{0z}$ respectively depend only on three non-dimensional parameters: the Prandtl number $\Pran = \nu / \kappa_T$, where $\nu$ is the kinematic viscosity and $\kappa_T$ is the thermal diffusivity, $\tau = \kappa_C / \kappa_T$, where $\kappa_C$ is the diffusivity of the solute, and the density ratio, $R_0 = \alpha T_{0z} / \beta C_{0z}$, where $\alpha$ and $\beta$ are derivatives of the equation of state:  $\alpha = - \rho^{-1} (\partial \rho /\partial T)$ and $\beta = \rho^{-1} (\partial \rho /\partial C)$. A linear fingering instability occurs provided $1 < R_0 < 1/\tau$, the lower limit being set by the requirement that the overall density be stably-stratified (whereas $R_0 < 1$ is convectively unstable), and the upper limit being the condition for marginal stability. 

Significant progress has been made to characterize mixing by fingering convection (in the absence of any other perturbing dynamics), using a combination of laboratory experiments \citep{turner1967,schmitt1979flux,linden1973structure,kunze2003ros}, theoretical models \citep{schmitt1979fgm}, and 3D direct numerical simulations \citep{stern2001sfu,yoshida2003numerical,Traxler2011a}. Recently, \citet{RadkoSmith2012} \citep[see also][]{Brownal2013} proposed a simple model in which the saturation of the fingering instability is attributed to the development of shear instabilities between upward- and downward-moving fingers. The model, once properly calibrated against direct numerical simulations, yields remarkably accurate estimates of the temperature and solute fluxes induced by fingering convection across much of parameter space. These flux laws can in principle be used to include the effects of small-scale fingering in global models (e.g. oceanic models or stellar evolution models). 

Further complications can arise in some regions of parameter space, however, where secondary large-scale instabilities are known to develop. One of the most well-known manifestations of such secondary instabilities is the formation of thermohaline staircases, which are often associated with salt fingering convection and have been observed in oceanic field measurements \citep{tait1968sot,tait1971ts,schmitt1987c, schmitt1995salt,schmitt2005enhanced}, laboratory experiments \citep{stern1969sfa,krishnamurti2003double, krishnamurti2009heat}, and more recently, numerical simulations \citep{radko2003mechanism,Stellmach2011}. 
\citet{radko2003mechanism} explained the formation of thermohaline staircases using a mean-field theory, where instability occurs through a positive feedback loop between horizontally-invariant variations in the local density ratio, and associated changes in the local turbulent fluxes of temperature and solute. \citet{radko2003mechanism} established that a necessary condition for the instability to occur is that $\gamma$, the ratio of the fingering temperature flux to the fingering solute flux, must be a decreasing function of the density ratio (hence the given name ``$\gamma-$instability"). This condition is satisfied in most geophysical fluids in a range of density ratios $1<R_0<R_{\rm crit}$ where $R_{\rm crit}$ depends on the fluid properties ($\Pran, \tau$). For salty water, for instance, $R_{\rm crit}$ is of the order of 4 \citep{Traxler2011a}, although the layering growth rate already drops significantly when $R_0 \simeq 2$. Radko's theory was quantitatively confirmed by 3D direct numerical simulations \citep{Stellmach2011}. 

Layering modes are not the only large-scale structures to emerge out of homogeneous fingering convection: as first discussed by \citet{stern1969cis} and \citet{holyer1981cis}, and later formalized by \citet{Traxler2011a}, fingering convection naturally excites large-scale internal gravity waves through a process called the collective instability. These waves typically grow more rapidly than the layering modes, and can exist in a region of parameter space that is stable to the $\gamma-$instability. They have been hypothesized to play a role in the formation of thermohaline staircases, either directly \citep{stern1969sfa} or indirectly \citep{Stellmach2011}. 

\subsection{Sedimentary fingering convection}
\label{sec:intro2}

Sedimentary fingering convection has, by contrast, received much less attention than standard fingering convection. Early experiments of \citet{houk1973descent} explored sedimentary fingering convection using particles such as taconite (iron) and clay (silicate) suspended in water. In an insulated beaker, warm sediment-laden water was held above cold and fresh water using a thin asbestos plate. When the plate was removed, sediment fingers were seen to form. This experiment was repeated for various particle concentrations, using either iron or clay, to measure finger descent rates. Non-dimensional descent rates were shown to scale with the buoyancy difference between the two layers, as in the case of standard fingering \citep{turner1967}, with a similar (perhaps slightly larger) coefficient of proportionality. The estimated downward sediment flux was also found to be consistent with the estimated salt flux in standard fingering convection \citep{turner1967}.  Similar conclusions were later reached by \citet{green1987importance}, looking at the fingers beneath buoyant sediment-laden gravity currents generated in lock release experiments: salt fingers and sediment fingers behave in qualitatively and often quantitatively similar ways.  

\citet{Hoyal1999a,Hoyal1999b} as well as \citet{Maxworthy1999} further analyzed the dynamics of sedimentary fingering using laboratory experiments. \citet{Maxworthy1999} investigated configurations similar to those of \citet{green1987importance}, using salt instead of heat as the stabilizing component. 
His experiments demonstrated that if the stabilizing density difference between the salty water in the lower layer and the fresh
water in the upper layer is comparable to the destabilizing density difference due to the sediment loading in the upper layer, the dynamics of the flow is strongly influenced by instabilities generated at the interface between the two fluids, as a result of the growth of a dense  ``nose" region of salty, particle-laden fluid along this interface. Recent linear stability and nonlinear computational investigations of this two-layer configuration by \citet{burns2012sediment,burns2015sediment} and \citet{Yual2013,Yual2014} yield further insight into the dynamics of double-diffusive sedimentation, which are often difficult to observe because of the opacity of the sediment field. In particular, the simulations by \citet{burns2015sediment} were able to clarify the mechanisms behind the  ``fingering" and  ``leaking" modes observed experimentally by \citet{Parsonsal2001}. 

\subsection{This work}
\label{sec:intro3}

The various findings discussed in the previous Section motivate the question of whether any of the other interesting phenomena observed in standard fingering convection, notably the formation of layers, also apply to sedimentary fingering convection. \citet{carazzo2013particle} first investigated layer formation in sedimentary fingering convection in the context of volcanic ash clouds. When a volcano erupts, a cloud of warm dust is launched into the air and continues to rise until it reaches a neutral level of buoyancy. The ash plume then spreads horizontally to form an umbrella cloud. By this time the large particles have fallen from the cloud, and only fine particles remain. Once the latter begin to settle, the cloud is observed to break up into a series of layers interspersed with particle-depleted bands. Inspired by these observations, \citet{carazzo2013particle} designed a  set of laboratory experiments that produce layering in sedimentary fingering convection. Starting with a tank filled with two well-defined fluid layers, of fresh water above salt water, they injected  particle-laden fresh water using a syringe into the bottom of the tank. Just as in the case of the volcanic ash, the particle laden water rose to a point of neutral buoyancy and then spread horizontally forming an umbrella cloud. As the particles began to settle, layers developed throughout the umbrella cloud. Whether similar layers develop in other experimental setups still remains to be determined, but there is already enough evidence of their existence to justify a theoretical investigation of layer formation in sedimentary fingering convection. 


In this work we shall therefore study various aspects of sedimentary fingering convection using a simple model developed in Section \ref{sec: The Model} in which the background temperature and sediment concentration gradients are assumed to be constant. This approach naturally complements the significant existing body of literature for initially layered systems discussed above. Section \ref{sec: linear stab} rapidly investigates the linear stability properties of the model, emphasizing the similarities and differences between sedimentary and standard fingering convection. In Section \ref{sec: basic flux} we compare the turbulent fluxes induced by sedimentary fingering convection to those measured in standard fingering convection. In Section \ref{sec: Extension of Radko 2003}, we then study the possibility of layer formation in sedimentary fingering convection using a model similar to the one originally developed by \citet{radko2003mechanism}, and validate our findings against 3D direct numerical simulations in Section \ref{sec: Comparing results to linear sim}. Finally, moving away from the constant background gradients model we then present 3D direct numerical simulations of layer formation in sedimentary fingering convection in Section \ref{sec: Gaussian}, in a setup where the mean concentration profile is free to evolve from any initial conditions. Our findings are then summarized and discussed in Section \ref{sec: Discussion}.

\section{The Model}\label{sec: The Model}

In this work we use a simple single-fluid model for the dynamics of the sediments, as in \citet{necker2002high} and \citet{burns2012sediment} for instance. This model assumes that the sediment particles are small spheres, and that their stopping time is much shorter that the typical turnover timescale of fluid motions. With this assumption, the velocity of the particles is equal to that of the fluid plus a net downward motion with terminal velocity 
\begin{equation}
V_s =  \frac{2r^2 (\rho_p - \rho_m) g }{9 \rho_m \nu} \, ,
\label{eq:vsed}
\end{equation}
where $\rho_m$ is the mean density of the fluid, $\rho_p$ is the solid density of the particles, $r$ is the radius of the particle, and $\nu$ is the kinematic viscosity. We also assume that the particles all have the same size, and that they do not collide or stick with each other. 


Within this formalism and otherwise using the Boussinesq approximation, the governing equations are 
\begin{eqnarray}\label{eq: SFC}
\label{eq: sed Boussinesq}
\frac{\p \bu}{\p t}+ \bu\cdot\nabla \bu=\frac{-\nabla p}{\rho_m} +\mathbf{g}\frac{ \rho}{\rho_m}+\nu \nabla^2  \bu,  \\
\frac{\p T}{\p t} + \bu\cdot\nabla T =\kappa_T\nabla^2 T,\\
\frac{\p C}{\p t} + (\bu-V_s \hat \be_z)\cdot \nabla C =\kappa_C \nabla^2 C,\\
\nabla \cdot \bu=0,
\end{eqnarray}
where $\bu$ is the fluid velocity, ${\bf g}$ is gravity, $T$ is the temperature perturbation away from a mean $T_m$, and $C$ is the sediment concentration.  The density perturbation $\rho$ away from the mean $\rho_m$ is related to $T$ and $C$ through a linearized equation of state,
\begin{equation}
\frac{\rho}{\rho_m} = - \alpha T + \beta C \, ,
\end{equation}
where $\alpha$ and $\beta$ were defined in Section \ref{sec:intro1}. Finally, the particles are assumed to have a small diffusivity $\kappa_C$ which can arise from random collisions with water molecules for instance (e.g. Brownian motion), or from long-range interactions mediated by the flow around the particles. 

In this Section (as well as in Sections \ref{sec: linear stab} to \ref{sec: Comparing results to linear sim}), the background we shall consider consists of an unbounded fluid with temperature and sediment concentration that both increase linearly with height. This setup is identical to the original setup of \citet{baines1969} in the thermohaline case. The fluid itself is motionless, and the background temperature $T_0(z)$ is expressed as
\begin{equation}
T_0(z) = T_m + T_{0z} z \, ,
\end{equation}
where $T_m$ is the mean temperature of the fluid, and $T_{0z}$ is the background temperature gradient, which is constant in time and space. The sediment concentration profile $C_0(z,t)$ on the other hand must be time-dependent to account for the sedimentation, and takes the form
\begin{equation} 
C_0(z,t) = C_{0z} (z + V_s t) \, .
\end{equation}
Note how the vertical concentration gradient $C_{0z} \equiv \partial C_0 / \partial z$ is constant even though $C_0(z,t)$ itself varies with time. It is easy to verify that $T_0(z)$ and $C_0(z,t)$ are solutions of the set of equations (\ref{eq: SFC}) when $\bu = 0$.   

Although this background state may seem overly simplistic and perhaps somewhat contrived, it is at least mathematically well-posed, and allows for a direct comparison of our results with the better-known case of fingering convection on linear gradients \citep{baines1969}. This is particularly important since \citet{radko2003mechanism} derived the $\gamma-$instability theory for layer formation in the thermohaline case within the same assumptions. As we shall demonstrate later, our model results can then be applied to explain, at least qualitatively, what happens in other cases where the initial concentration profile is not linear (see Section \ref{sec: Gaussian}). 

Next we consider perturbations $\tilde T$ and $\tilde C$ to this background state: 
\begin{eqnarray}
\label{eq: linearize}
T=T_0(z)+\tilde T \, ,\\
C=C_0(z,t)+\tilde C \, .
    \end{eqnarray}
The evolution equations for the perturbations are 
\begin{eqnarray}
\frac{\p \bu}{\p t}+  \bu\cdot\nabla \bu= -\nabla \tilde p +(\alpha \tilde T - \beta \tilde C)g \be_z+\nu \nabla^2  \bu, \label{eq: sed Boussinesq2}\\
\frac{\p \tilde T}{\p t} + \bu\cdot\nabla \tilde T + w T_{0z} =\kappa_T\nabla^2 \tilde T,\\
\frac{\p \tilde C}{\p t} + \bu\cdot \nabla \tilde C + w C_{0z} -V_s\frac{\p\tilde C}{\p z}=\kappa_C \nabla^2 \tilde C,  \\
\nabla \cdot  \bu=0,\label{eq: sed Boussinesq3}
    \end{eqnarray}
where $\tilde p$ is the pressure perturbation away from hydrostatic equilibrium (in the background state) divided by $\rho_m$. 
To eliminate any influence from the boundaries, we employ triply-periodic boundary conditions for all perturbations (e.g. $\tilde T$, $\tilde C$, $\tilde p$ and $\bu$). 

We non-dimensionalize the equations as in \citet{radko2003mechanism}, using the expected finger scale, 
\begin{equation}
\label{eq:length d}
d=\left(\frac{\kappa_T \nu}{g\alpha T_{0z}}\right)^{1/4} \, ,
\end{equation}
as unit length. We then define the unit time $[t]$, the unit velocity $[u]$, the unit temperature $[T]$ and the unit concentration $[C]$ as 
\begin{eqnarray}
[t]=\frac{d^2}{\kappa_T},  [u]=\frac{\kappa_T}{d},
 [T]=dT_{0z},  [C]=\frac{\alpha}{\beta}dT_{0z}.
    \end{eqnarray}\label{eq: characteristics}
The non-dimensional form of equations (\ref{eq: sed Boussinesq2})-(\ref{eq: sed Boussinesq3}) is 
\begin{eqnarray}
\label{eq:non_dimu}
\frac{1}{\Pr}\left(\frac{\partial \bu}{\partial t}+\bu\cdot\nabla\bu\right)=-\nabla p +(T-C)\hat e_z+\nabla^2\bu,\\
\label{eq:non_dimT}
\frac{\partial T}{\partial t}+\bu\cdot\nabla T+w=\nabla^2T,\\
\label{eq:non_dimC}
\frac{\partial C}{\partial t}+\bu\cdot\nabla C +\frac{w}{R_0} -V \frac{\p C}{\p z}=\tau\nabla^2C,\\
\label{eq:non_dimm}
\nabla \cdot \bu=0,
\end{eqnarray}
where the tildes are henceforth dropped. The non-dimensionalization results in four dimensionless parameters: the Prandtl number $\Pran=\nu/\kappa_T$, the diffusivity ratio $\tau=\kappa_C/\kappa_T$, and the density ratio $R_0=\alpha T_{0z}/\beta C_{0z}$, which are the usual parameters describing standard fingering convection (see Section \ref{sec:intro1}), together with the non-dimensional settling velocity $V=V_s d/\kappa _T$, which is specific to the sedimentary problem.

The Prandtl number of water decreases from about 10 for nearly freezing water, down to about 1 for water near the boiling point, with the standard value of $7$ for water at 20$^\circ$C. A typical value of $\tau$ for the heat/salt system is about 1/100. Estimates of $\tau$ for sedimentary systems can be obtained from the sum of the contributions from Brownian motion and from hydrodynamic diffusion of the individual sediment grains due to their long-range hydrodynamic interactions. For the latter, several authors \citep{ham1988hindered,nicolai1995particle,segre2001effective} find experimental values of the particle diffusivity on the order of $r V_s$. This gives 
\begin{equation}
\tau  = \frac{k_{\rm B}T_m}{6\pi \rho_m \nu r \kappa_T} + \zeta \frac{rV_s}{\kappa_T} =  \frac{k_{\rm B}T_m}{6\pi \rho_m \nu r \kappa_T} + \zeta \frac{r}{d} V \mbox{  .}
\end{equation}
where $k_{\rm B}$ is the Boltzmann constant,  and $\zeta$ is a factor of order unity that depends on the volume fraction, e.g. \citet{segre2001effective}.

Finally, in order to estimate $V$ we first need to estimate $d$ (see equation \ref{eq:length d}). Assuming a typical value of $T_{0z}$ in the mid-latitude thermocline of the ocean to be about $0.01 ^\circ C$ m$^{-1},$ as well as $\alpha = 2 \times 10^{-4}$K$^{-1}$, $\kappa_T = 0.145\times 10^{-6}$m$^2/$s, $\nu = 10^{-6}$m$^2/$s which are appropriate for water at room temperature, the dimensional value for the lengthscale is $d\sim 2.7$ cm. Using equation (\ref{eq:vsed}), we find that for particles with typical densities of around 2,000 kg/m$^3$ (e.g. clays for instance) settling in water at ambient temperature,
\begin{equation}
 V = \frac{V_s d}{\kappa_T} \simeq  0.45 \left( \frac{r  }{ 10 \mu m } \right)^2  \, .
\end{equation}

\begin{figure}
\centering
\includegraphics[width=0.5\linewidth]{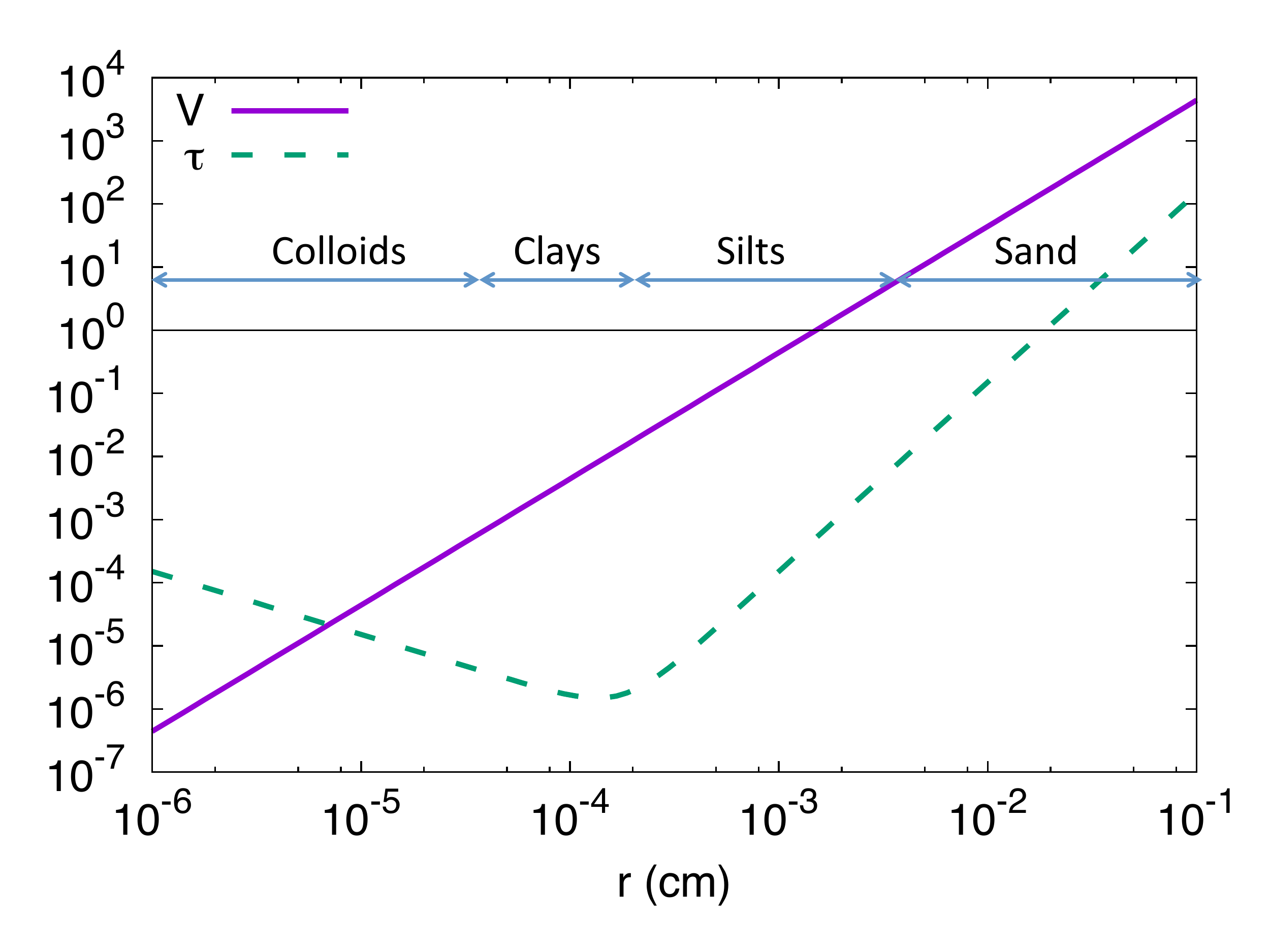}
\caption{Variation of the non-dimensional settling velocity $V$ and of the diffusivity ratio $\tau$ for varying particle radius; see main text for details. For low values of $r$, $\tau$ is dominated by the contribution from Brownian motion, while for larger values of $r$, $\tau$ is dominated by hydrodynamic diffusion coming from long-range interactions.  }\label{fig:nondim}
\end{figure}

Figure \ref{fig:nondim} shows estimates of $V$ and $\tau$ for particles of varying radii $r$, ranging from the smallest colloids to the coarser sand, with $T_{0z} = 0.01 ^\circ C$ m$^{-1}$, $\alpha = 2 \times 10^{-4}$K$^{-1}$, $\kappa_T = 0.145\times 10^{-6}$m$^2/$s, $\nu = 10^{-6}$m$^2/$s, $\rho_p = 2,000$ kg/m$^3$ and $\zeta = 1$. We see that $\tau$ is of order $10^{-4}$ for clay particles, and grows rapidly with particle size. It is interesting to see that for large enough particle sizes (e.g. for coarser sand), $\tau$ could even be larger than one (a limit we do not investigate in this paper).  

\section{Linear Stability}\label{sec: linear stab}

\begin{figure}
\centering
\includegraphics[width=0.5\linewidth]{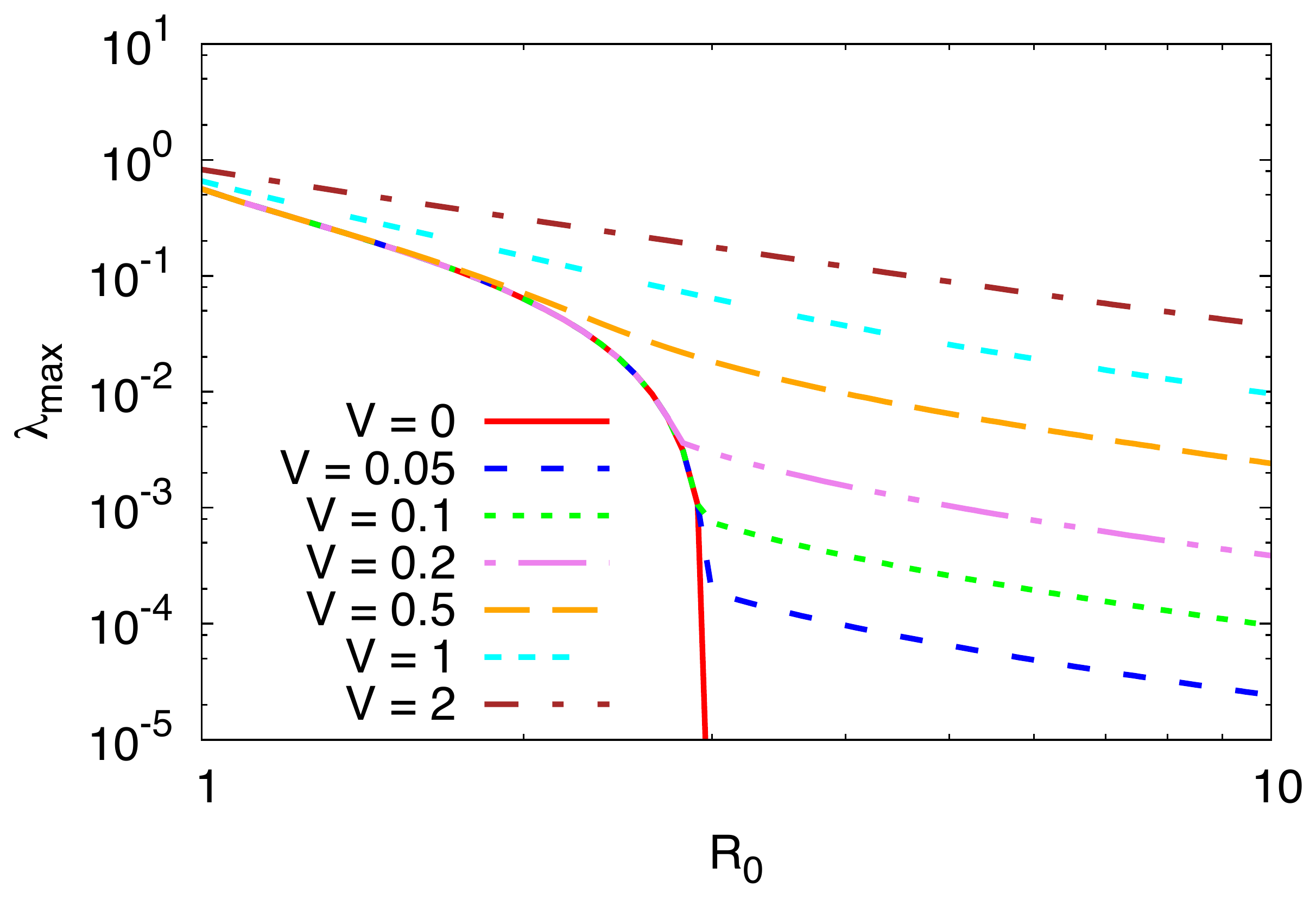}
\caption{Variation of the growth rate $\lambda_{\rm max}$ of the fastest-growing mode with $R_0$ and $V$, for $\Pran = 7$ and $\tau = 1/3$.  \label{fig:lineartheory}}
\end{figure}

We first investigate the linear stability of sedimentary fingering convection with constant background gradients (see Alsinan et al., {\it submitted}, for a more comprehensive study of this problem). 
Assuming normal modes of the kind 
\[(u,v,w,p,T,C)=(\hat{u},\hat{v},\hat{w},\hat{p}, \hat{T}, \hat{C})\exp(ilx+imy + ikz+\lambda t)\]
and substituting them into the linearized version of the non-dimensional equations (\ref{eq:non_dimu})-(\ref{eq:non_dimm}) we obtain a cubic equation for the growth rate $\lambda$, as $\lambda^3 + a \lambda^2 + b \lambda + c = 0$ where 
\begin{eqnarray}
&& a = (1+\Pr+\tau)K^2 -ikV \, ,\\
&& b =  (\Pr + \tau + \Pr \tau)K^4 - ikK^2 (\Pr + 1) V + \frac{l^2+m^2}{K^2} \Pr (1 - R_0^{-1})  \, ,  \\ 
&& c =  \Pr \tau K^6 - ik  K^4 \Pr V +  \frac{l^2+m^2}{K^2} \Pr \left[ (\tau  - R_0^{-1})K^2 - ikV   \right]   \, ,
\end{eqnarray}
where $K^2 = l^2 + m^2 + k^2$. This recovers the theory of \citet{baines1969} when $V = 0$. Since the system is horizontally isotropic, $\lambda$ only depends on the total horizontal wavenumber $k_h = \sqrt{l^2 + m^2}$. 
To find the growth rate $\lambda_{\rm max}$ of the most unstable mode for given parameter values $R_0$, $\Pran$, $\tau$ and $V$, we maximize the real parts of the solutions of the cubic over all possible $k_h$ and $k$. The results are presented in Figure \ref{fig:lineartheory}, for $\Pran = 7$, $\tau = 1/3$, and varying $R_0$ and $V$. Note that this is a very moderate value of $\tau$, for simplicity, and for comparison with the results of the direct numerical simulations presented later. However, these basic findings hold for $\tau \ll 1$ as well (see Alsinan et al., {\it submitted}). 

When $V = 0$, the growing modes are always direct modes (i.e. $\lambda$ is real), the fastest-growing modes have $k = 0$ if the domain is infinitely tall \citep{radko2013double}, and, as discussed in Section \ref{sec:intro1}, unstable modes only exist if $R_0  < 1/\tau$. When $V \neq 0$, on the other hand, these statements do not necessarily remain true. 

For small $V$, namely $0 < V \ll 1$, we see that $\lambda_{\rm max}$ is essentially unaffected by settling for most values of $R_0 \in [1,1/\tau]$. This is not surprising, since the terms that contain $V$ in the cubic equation are always multiplied by $k$. As a result, as long as the fastest-growing mode has $k = 0$, its dynamics according to linear theory are independent of $V$. However, we see that for $R_0$ approaching $1/\tau$,  $\lambda_{\rm max}$ begins to depend on $V$ which implies that $k = 0$ is no longer the fastest-growing mode.  For larger values of $V$, this is in fact always the case: the fastest-growing mode always has a vertical wavenumber $k \ne 0$ and $\lambda_{\rm max}$ always exceeds its corresponding non-sedimentary value. It also acquires an imaginary part, associated with the downward vertical translation of the sediment field. The reason for this transition between two fundamentally different kinds of modes (from $k = 0$ to $k \ne 0$) is clarified by Alsinan et al. ({\it submitted}).

Finally, we also see that unstable modes now exist for arbitrarily large values of  $R_0$ as long as $V > 0$. As $R_0$ increases, the horizontal wavenumber $k_h$ of the fastest-growing mode decreases. If the horizontal extent of the domain is limited, then there exists a maximum value of $R_0$ for instability. However if the domain is horizontally infinite, then one can always find a value of $k_h$ small enough for instability. Note, however, that the region of parameter space unstable to sedimentary fingering convection where $R_0 > 1/\tau$ may only be of academic interest, since $\tau$ is usually very small -- hence $1/\tau$ is very large  -- for standard sediment sizes. 

\section{Nonlinear saturation of the sedimentary instability}\label{sec: basic flux}

We now look at the nonlinear development and saturation of the sedimentary fingering instability in the setup considered, i.e. for fixed background gradients of temperature and sediment concentration. To do so, we use the PADDI code \citep{Traxler2011a,Stellmach2011}, having modified it to incorporate sedimentation. In this Section, unless specifically mentioned, we solve equations (\ref{eq:non_dimu})-(\ref{eq:non_dimm}) in a triply-periodic domain, with $\Pran=7$ and $\tau=1/3$, for various density ratios and various non-dimensional settling velocities (see Tables 1-3). This value of $\Pran$ is appropriate to model water. The value of $\tau$ chosen is significantly larger than what one would expect in any physically realistic problem. Indeed, as shown in Figure \ref{fig:nondim}, for clays and silts $\tau \ll 1$, ranging from $10^{-6}$ to $10^{-2}$. It is only for the fine sands ($r \sim 0.1$mm) that $\tau \sim 0.1-1$, but for these particle sizes $V \gg 1$. Unfortunately, the time needed to run the simulations would be prohibitive for smaller values of $\tau$ (due to the high resolution required), or for larger values of $V$ (due to the smaller timestep required), especially considering that our goal is to explore the range of possible dynamical behaviors of sedimentary fingering convection arising from varying both $R_0$ and $V$.

\begin{table}
\begin{center}
\begin{tabular}{ccccccccc}
A: & $R_0 = 1.1 $ &   &   &    &   &   &   &    \\
$V$  & $-\langle wT \rangle_t$ & $-\langle wC \rangle_t$ & $L_x\times L_y \times L_z$ &  $N_x \times N_y \times N_z$ & L? & H & $\Lambda_{\rm obs}$ & $\Lambda_{\rm theor}$  \\
$0^{(a)}$ & 185.13$\pm$6.67 & 223.23$\pm$10.32 & $83 \times 83\times268$  &  not avail.  & N &  &  &  \\ 
$0^{(a)}$ & not avail.  & not avail.  & $ 335\times 335\times536$    &  $ 576\times 576\times1152$   & Y & 268 & 0.01 & 0.0135 \\ 
0.1 & 189.55$\pm$9.38 & 	226.61$\pm$10.07	& $100\times100\times100$ & $384\times 384\times384$ & N	&  &  &   \\
0.1 & 183.58$\pm$16.33 &220.15$\pm$15.90& $200\times25\times200$& $768\times96\times768$ & Y & 200& 0.017 & 0.022	\\ 
0.5 &  192.76$\pm$12.12	 & 232.70$\pm$12.96  & $100\times100\times100$    & $384\times 384\times384$   & N	&  &  &   \\
1$^{(b)}$ & 218.67$\pm$13.57  & 269.47$\pm$14.32  & $100\times100\times100$   &  $192\times 192\times192$  &  N	&  &   &    \\
1 & 228.61$\pm$19.82 & 280.46$\pm$21.47 &	$100\times100\times100$  & 	$384\times 384\times384$ &   N	& &   &       \\
2 &  	273.49$\pm$24.76  &353.68$\pm$26.26   & $100\times100\times100$   &  $384\times 384\times384$  &  N	&  &   &    \\
\\
B: & $R_0 = 1.3 $ &   &   &    &   &   &   &    \\
$V$  		& $-\langle wT \rangle_t$ & $-\langle wC \rangle_t$		& $L_x\times L_y \times L_z$ 	&  $N_x \times N_y \times N_z$ & L? & H & $\Lambda_{\rm obs}$ & $\Lambda_{\rm theor}$  \\
$0^{(a)}$ 	& 73.95$\pm$1.65 	& 73.95$\pm$1.65  	& $83 \times 83\times268$  	&  not avail.  			& N &  &  &  \\ 
0.1$^{(d)}$ & 77.26$\pm$2.35 & 95.85$\pm$2.96	& $100\times100\times100$ & 	$192\times192\times192$ & 	N&  &  & 	\\		
0.1 &	74.05$ \pm$1.93 & 92.01$\pm$2.21	& $100\times100\times200$ & 	$192\times192\times384$ & N&  &  &  \\
0.5$^{(b)}$& 69.82$ \pm$2.75 & 87.66$ \pm$3.12 & 	$100\times100\times100$ & 	$192\times192\times192$ & 	N& & &  \\
0.5 & 70.30$\pm$2.76 & 88.32$\pm$3.13 & $100\times20\times400$ & $384\times72\times1536$ & Y  & 400  & 0.0045  & 0.0021 \\ 
1$^{(b,c)}$ & 72.89$ \pm$5.09 & 94.74$\pm$5.32	& $100\times100\times100$ & 	$192\times192\times192$ &	Y	& 100	& 0.011 &	0.02 \\
1$^{(b,c)}$& 69.33$ \pm$2.60 &	90.58$ \pm$2.63 &	$100\times100\times200$ & 	$192\times192\times384$ & 	Y	& 100	& 0.011 &	0.02 \\
\\
C: & $R_0 = 1.5 $ &   &   &    &   &   &   &    \\
$V$  & $-\langle wT \rangle_t$ & $-\langle wC \rangle_t$& $L_x\times L_y \times L_z$ &  $N_x \times N_y \times N_z$ & L? & H & $\Lambda_{\rm obs}$ & $\Lambda_{\rm theor}$  \\
0$^{(a)}$ & 41.87$\pm$0.99  &   52.86$\pm$1.16	 &$83 \times 83\times268$  	&  not avail. 			& N &  &  &  \\ 
0.05 &	44.15$\pm$1.16 &	55.47$\pm$1.35& $100\times100\times100$ & 	$192\times192\times192$ & 	N&  & &  \\
0.1 & 43.8$\pm$1.45 &55.03$\pm$1.67	 &$100\times100\times100$ & 	$192\times192\times192$ & 	N& &  &  \\
0.25 & 	42.68$\pm$1.73 &	53.71$\pm$2.02&$100\times100\times100$ & 	$192\times192\times192$ & 	N&  &  &  \\
0.25 &	40.78$\pm$1.07 &	51.54$\pm$1.24&$100\times100\times200$ & 	$192\times192\times384$ & 	N&  &  &  \\
0.5 & 36.55$\pm$0.96 &	46.60$\pm$1.12&$100\times100\times100$ & 	$192\times192\times192$ & 	N&  &  &  \\
0.5 &  35.60$\pm$1.28 & 45.56$\pm$1.50 & $100 \times 20 \times 400$ & $192\times36 \times768$ & Y & 200 & 0.0033 & 0.0005  \\ 
0.5 & 20.57$\pm$1.26 & 26.23$\pm$1.53 & $256 \times 0 \times 512$ & $1024 \times 0 \times 2048$ & Y & 128 & 0.004 & N/A \\
1$^{(b)}$ & 26.98$\pm$0.89& 36.6$\pm$1.00 &$100\times100\times100$ & 	$192\times192\times192$ &	Y	&100 &	0.012 & 0.0067 \\
2$^{(b,c)}$ & 48.48$\pm$1.61 &	72.35$\pm$1.70 & $100\times100\times100$ &	$192\times192\times192$ &	Y &	100	 & 0.05 &	0.021 \\
2$^{(c)}$ & 47.48$\pm$2.86 & 70.88$\pm$3.33 & $100\times100\times100$ &	$384\times 384\times384$ &	Y &	100	 & 0.05 &	0.021  \\
\\
\end{tabular}
\end{center}
\caption{\small Summary of our numerical experiments at $\Pr = 7$ and $\tau = 1/3$. In each set, the density ratio is held constant at the value indicated. The first three column show $V$, $-\langle wT\rangle_t$ and $-\langle wC \rangle_t$ ($\pm$ the r.m.s. fluctuations around the mean), where $\langle \cdot \rangle$ denotes a volume average, and the subscript $t$ denotes a time-average. Note that the fluxes are measured in the homogeneously turbulent phase prior to the formation of layers (if they do form). The fourth column shows the domain size. The fifth column shows the effective number of meshpoints in each direction. The sixth column indicates whether layers form or not during the simulation. The seventh column shows the layer height in the initial staircase. The eighth column is the measured growth rate of the dominant layering mode (whose wavelength is the same as the initial layer height) and finally the last column shows the theoretical growth rate of that mode (according to the theory of Section \ref{sec: Extension of Radko 2003}).  } 
\begin{tablenotes}
        \item [] (a) data from \citet{Stellmach2011}
        \item [] (b) runs slightly under-resolved 
        \item [] (c) runs slightly under-resolved after layer formation
        \item [] (d) coherent streams 
      \end{tablenotes}
\end{table}

\begin{table}
\begin{center}
\begin{tabular}{ccccccccc}
E: & $R_0 = 1.7 $ &   &   &    &   &   &   &    \\
$V$  & $-\langle wT \rangle_t$ & $-\langle wC \rangle_t$& $L_x\times L_y \times L_z$ &  $N_x \times N_y \times N_z$ & L? & H & $\Lambda_{\rm obs}$ & $\Lambda_{\rm theor}$  \\
0$^{(a)}$  &19.28$\pm$0.40 & 24.76$\pm$0.48	&$83 \times 83\times268$  	&  not avail.  			& N &  & &  \\ 
0.1 & 20.09$\pm$0.55 &	25.79$\pm$0.66&$100\times100\times100$ & 	$192\times 192\times192$ &	N& &  & \\
0.5$^{(d)}$  & 23.02$\pm$1.07 &29.14$\pm$1.26 &$100\times100\times100$ & 	$192\times 192\times192$ &	N&  & &  \\
0.5 & 19.75$\pm$0.78 & 25.36$\pm$0.93 & $100 \times 20 \times 400$ & $192\times 36 \times 768$ & Y & 200 & 0.0017 &  stable  \\ 
0.5 & 7.65$\pm$0.53 & 9.96$\pm$0.68 & $512 \times 0 \times 1024$ & $2048 \times 0 \times 4096$  & Y & 102 & 0.0077 & N/A \\
1$^{(c,d)}$	 & 10.63$\pm$0.26 &14.85$\pm$0.35 &$100\times100\times100$ & 	$192\times 192\times192$ &	Y& 100 &	0.022 & 0.0036 \\ 
1$^{(c)}$ & 	10.27$\pm$0.21 &	14.38$\pm$0.29 &$200\times100\times 200$ & 	$384\times 192\times384$ 	& Y	& 66 &  0.019 &  stable \\
\\
F: & $R_0 = 2.0 $ &   &   &    &   &   &   &    \\
$V$  & $-\langle wT \rangle_t$ & $-\langle wC \rangle_t$ & $L_x\times L_y \times L_z$ &  $N_x \times N_y \times N_z$ & L? & H & $\Lambda_{\rm obs}$ & $\Lambda_{\rm theor}$  \\
0$^{(a)}$ &7.56$\pm$0.16 &   9.54$\pm$0.19	&$83 \times 83\times268$  	&  not avail.  			& N &  &  &  \\ 
0.1 & 7.40$\pm$0.26 & 	9.35$\pm$0.32 &$100\times100\times100$ & 	$192\times 192\times192$ &	N&  &  &  \\
0.25 & 7.468$\pm$0.26	& 9.42$\pm$0.31 &$100\times100\times100$ & 	$192\times 192\times192$ &	N& &  &  \\
0.5$^{(d)}$& 7.99$\pm$0.46 &	9.99$\pm$0.54 &$100\times100\times100$ & 	$192\times 192\times192$ &	N& &  &  \\
0.5 & 7.92$\pm$0.29 &9.94$\pm$0.35 & $100\times100\times200$ & 	$192\times 192\times384$ &	Y& 200 & 0.0013 & stable \\
0.75 &2.53$\pm$0.11 &3.41$\pm$0.15&$100\times100\times100$ & 	$192\times 192\times192$ &	Y	 & 100	& 0.014	&  stable \\
1$^{(c)}$ & 2.16$\pm$0.03 & 3.16$\pm$0.04 &$100\times100\times100$ & 	$192\times 192\times192$ & Y &	33 & 	0.025 &  0.0083 \\
1.5$^{(c)}$ & 	5.49$\pm$0.22 & 8.95$\pm$0.28 & $100\times100\times100$ & 	$192\times 192\times192$ & Y &  33& 0.05&  stable  \\
2$^{(b,c)}$	&10.53$\pm$0.47 & 18.21$\pm$0.57 & $100\times100\times100$ & 	$192\times 192\times192$ & Y  & 50	& 0.055 &  0.026 \\
2 &	10.44$\pm$0.41 & 18.12$\pm$0.55	& $100\times100\times100$ & 	$384\times 384\times384$ & Y &	50& 0.06	& 0.026 \\
2$^{(b,c)}$&	10.35$\pm$0.33&18.02$\pm$0.52 & $200\times100\times200$ & 	$384\times 192\times384$ & Y  & 50 & 0.055 & 0.026\\
2 &	10.50$\pm$0.31&18.35$\pm$0.48& $100\times100\times200$ & 	$384\times 384\times768$ & Y  & 50 & 0.06 & 0.026\\
\\
G : & $R_0 = 2.5 $ &   &   &    &   &   &   &    \\
$V$  & $-\langle wT \rangle_t$ & $-\langle wC \rangle_t$ & $L_x\times L_y \times L_z$ &  $N_x \times N_y \times N_z$ & L? & H & $\Lambda_{\rm obs}$ & $\Lambda_{\rm theor}$  \\
0.05$^{(d)}$ & 2.18$\pm$0.26 & 	2.49$\pm$0.3 & $100\times100\times100$ &$192\times 192\times192$ & N &  &  &  \\  
0.05 & 1.62$\pm$0.06 & 	1.86$\pm$0.06 & $100\times100\times200$ &$96\times 96\times192$ & N & &  &  \\  
0.1$^{(d)}$ &2.23$\pm$0.31  &2.55$\pm$0.35 &  $100\times100\times100$ &  $96\times 96\times96$ &  N & & & \\
0.1 & 1.57$\pm$0.05 &1.80$\pm$0.055 &  $100\times100\times200$ &  $96\times 96\times192$ & N &  & &   \\
0.25$^{(d)}$ & 3.22$\pm$0.9&3.65$\pm$1.00 &  $100\times100\times100$ &  $96\times 96\times96$ & N &  &  & \\ 
0.25	& 1.39$\pm$0.04 &	1.59$\pm$0.04	&  $200\times100\times200$ &  $192\times 96\times192$ & Y	& 200 &	0.00014	&   0.0009\\
0.5&	0.27$\pm$0.02   &	0.32$\pm$0.02	&  $100\times100\times100$ &$192\times 192\times192$ &	Y	& 100&	0.0027	& 0.031 \\ 
0.75 	& 0.18$\pm$0.003 & 0.24$\pm$0.005 &  $100\times100\times100$ &  $96\times 96\times96$ &	N &  & &  \\ 
1 & 	0.42$\pm$0.02 &  0.66$\pm$0.04 & $100\times100\times100$ &$192\times 192\times192$ & Y	& 100&	0.0036   & 0.012 \\
1.5$^{(c)}$ & 1.47$\pm$0.07 &2.71$\pm$0.14 &   $100\times100\times100$ &  $96\times 96\times96$ & Y	& 33 & 0.026 & 0.0077 \\
2 & 3.90$\pm$0.14 &7.79$\pm$0.23 & $100\times100\times100$ &$192\times 192\times192$ & Y & 	50	& 0.028  & 0.017 \\
\end{tabular}
\end{center}
\caption{(continued from Table 1) } 
\end{table}

\subsection{Typical results}

\begin{figure}
\centering
\includegraphics[width=0.7\linewidth]{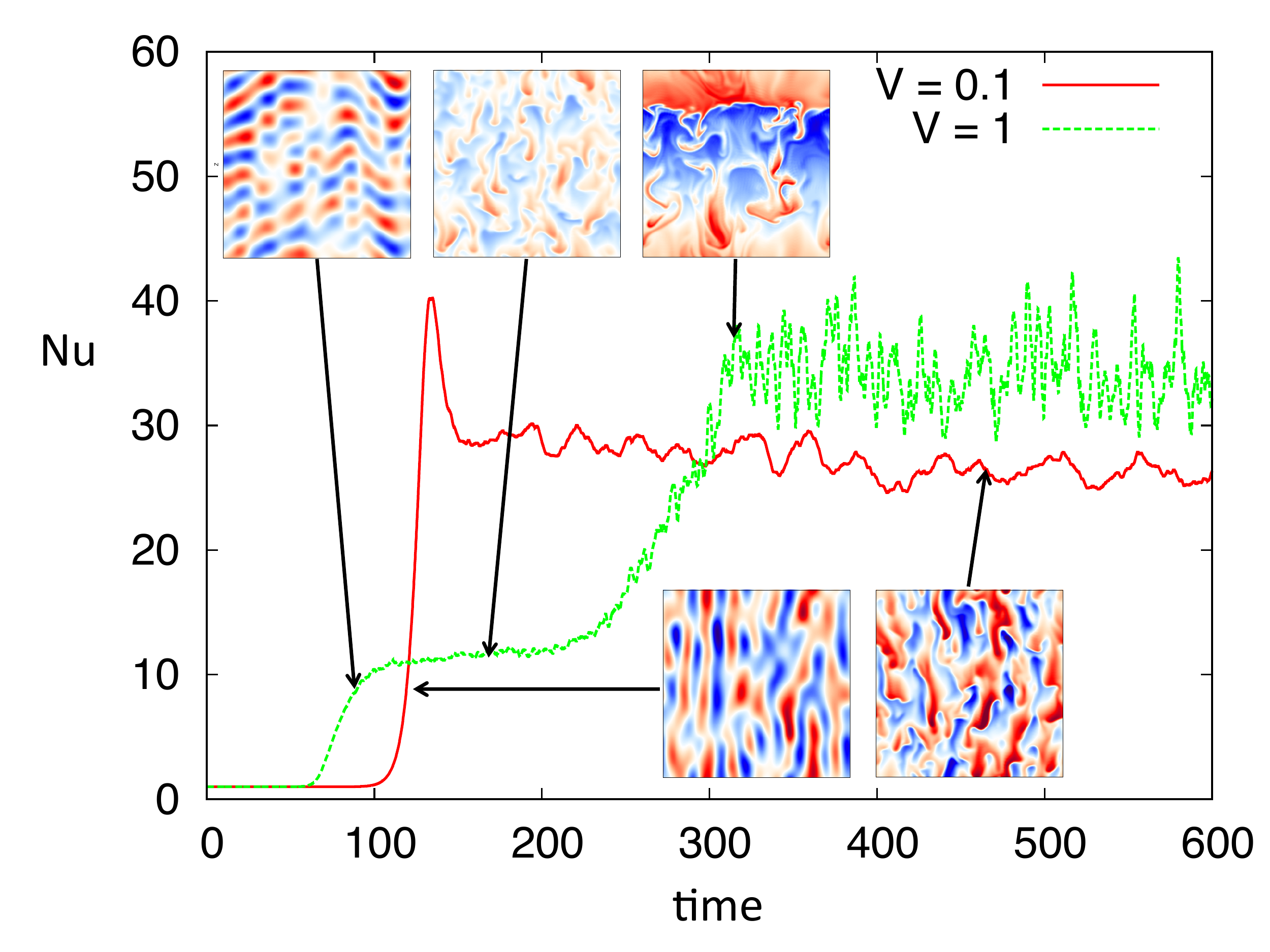}
\caption{Variation of the Nusselt number $\Nu(t) = 1 - \langle w T \rangle$ (where $\langle \cdot \rangle$ denotes a volume average over the entire domain) for two simulations with $R_0 = 1.7$, $\Pran = 7$ and $\tau = 1/3$, with $V = 0.1$ and $V = 1$ respectively, in a domain of size $100\times100\times100$. Shown in the insets are snapshots of the corresponding sediment concentration field at selected times. The color scales are different in each snapshot, to emphasize the dynamics at play. \label{fig:figuretypes}}
\end{figure}

Figure \ref{fig:figuretypes} contrasts the evolution of two simulations, one that has a small settling velocity ($V = 0.1$) and one with a larger one ($V = 1$), for $R_0 = 1.7$ which is roughly mid-way through the standard fingering instability range at these parameters. In both cases the domain size is $100\times100\times100$ (in units of $d$). 

For $V = 0.1$, the behavior of the system is qualitatively identical to what one would expect in the non-sedimentary case (at least, in this domain size, see below for more). The dominant modes of linear instability are vertically invariant tubes, that later become unstable to shear and saturate into a state of homogeneous fingering convection. No layers are observed to form during the simulation timeframe at these parameter values, which is expected since the $\gamma-$instability is not present at this density ratio \citep[see][as well as Figure \ref{fig:gamma}]{Traxler2011a,Stellmach2011}. 

For $V = 1$, notable differences emerge. The early behavior of the instability shows that the dominant modes are no longer vertically invariant, but instead inclined, and gradually propagate downward with time. The instability growth rate is also larger at $V =1$ than at $V = 0.1$. This is consistent with the findings of Section \ref{sec: linear stab} at these parameter values. 
Interestingly, the amplitude of the non-dimensional turbulent heat flux at saturation for $V = 1$ is smaller than at $V = 0.1$, a result we attribute to the fact that vertical motion is no longer preferred for larger $V$. Even more interestingly, we see a transition from homogeneous to layered sedimentary fingering convection. As in the case of layer formation in normal fingering convection, this transition is associated with an increase in the turbulent fluxes through the system. The formation of layers in sedimentary fingering convection is studied in Section \ref{sec: Extension of Radko 2003}. 

\subsection{Systematic exploration of parameter space}
\label{sec:basicfluxes}

Tables 1 and 2 (see also Table 3) summarize the results of our numerical experiments, which cover a range of density ratios $R_0$ and settling velocities $V$. In each experimental set (A through G) $R_0$ is held fixed while $V$ is varied. The domain size and resolution used for each run is provided. In each case, we report on the non-dimensional mean turbulent temperature and sediment fluxes $\langle wT \rangle_t$ and $\langle wC\rangle_t$ (where $\langle \cdot \rangle$ denotes a volume average, and the subscript $t$ denotes a time-average), as measured in the homogeneously turbulent phase prior to the formation of layers if applicable. We also report on whether a staircase forms or not (within the timeframe of the simulation), and how tall the layers initially are. The other columns are discussed in Section \ref{sec: Comparing results to linear sim}.

\begin{figure}
\centering
\includegraphics[width=\linewidth]{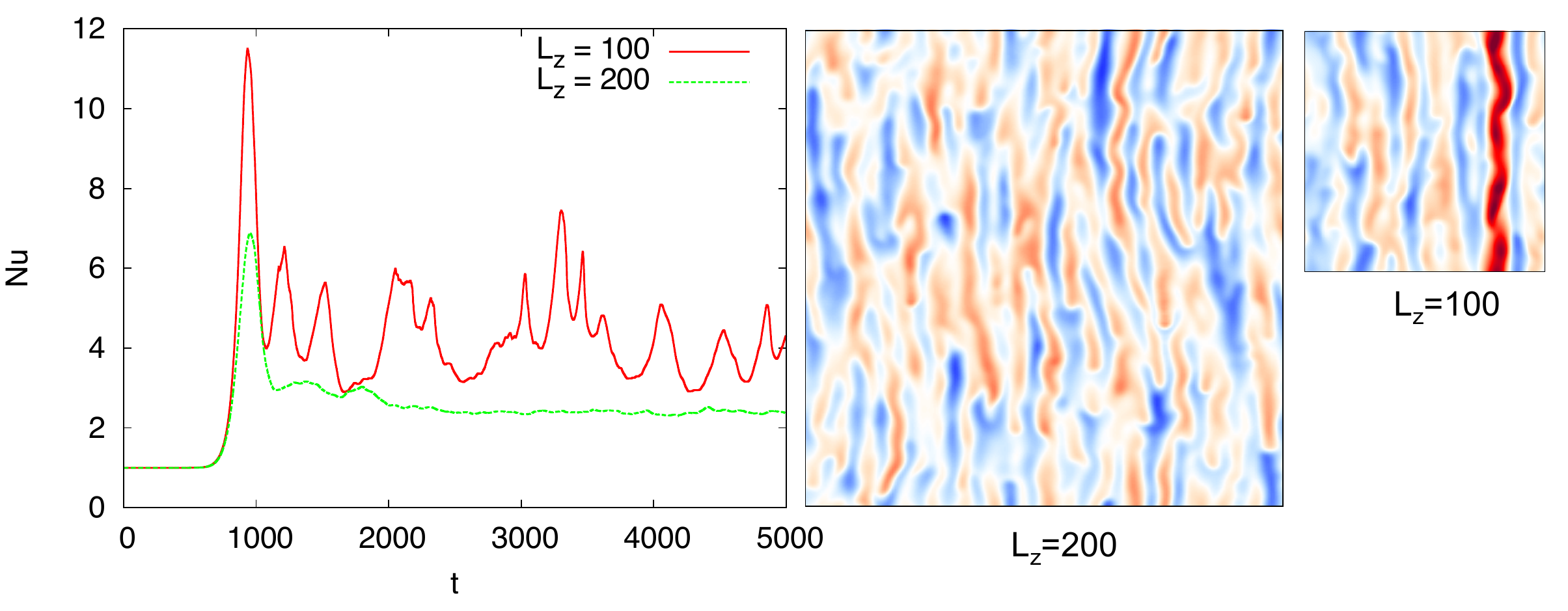}
\caption{Variation of $\Nu(t)  = 1  - \langle w T\rangle$ with time for two simulations with $R_0 = 2.5$, $\Pran = 7$, $\tau = 1/3$, and $V = 0.25$ in domain sizes with heights $L_z = 100d$ and $L_z = 200d$ respectively. At these parameter values the measured fluxes in the two different domains vary significantly. This can be attributed to the artificial formation of vertically invariant streams in which fluid rapidly moves up or down, shown in the snapshots. Both snapshots show the sediment concentration field with the same color scale. These streams appear in the small-domain simulations, but do not in the larger-domain one. \label{fig:weird}}
\end{figure}

Note that the domain size and resolution change depending on the selected values of $R_0$ and $V$. In the case of non-sedimentary fingering convection, it is well-known that the Reynolds number of the turbulent state decreases as $R_0$ increases. For this reason, we use a larger resolution at low density ratios than at high density ratios. However, we found that the resolution also had to be increased with increasing $V$ at fixed $R_0$. In a few cases, we report on the results of simulations that are visually slightly under-resolved. The measured fluxes in these simulations do not differ from the higher-resolution ones by a factor larger than their intrinsic variability, however (see for instance the cases with $R_0 = 1.1$ and $V = 1$, or $R_0 = 2$ and $V = 2$). Also note that for most simulations, a domain of size $100 \times 100 \times 100$ was sufficient to get consistent measurements of the fluxes: doubling the size of the domain did not significantly change $\langle wT \rangle_t$ and $\langle wC\rangle_t$ (see for instance the cases with $R_0 = 1.3$ with $V = 0.5$, and $R_0= 1.5$ with $V= 0.25$). However, this was not true for $R_0 = 2.5$ and $0 \le V < 0.5$. In that parameter range, the fingers are naturally very tall and laminar; vertically invariant streams emerge and dominate turbulent heat and particle transport, driving significantly higher values of $\langle wT \rangle_t$ and $\langle wC\rangle_t$.  These streams disappear, however, when the domain size is doubled (see Figure \ref{fig:weird}). 

Finally, we have run a few two-dimensional (2D) simulations in large domains at high resolution, to test the difference between 2D and 3D runs (see for instance the cases with $R_0 = 1.5$, $V = 0.5$ and $R_0 = 1.7$, $V = 0.5$), as well as 3D simulations in tall, thin domains (with $L_y = 20$, which corresponds to about two finger widths). We find that, while the fluxes measured in 3D thin domains are quite similar to the ones obtained in thicker ones, those measured in the 2D runs are substantially smaller than in 3D. Hence, 2D simulations should not be used to get quantitatively accurate estimates of fingering fluxes, but are adequate for qualitative studies \citep[see also][]{stern2001sfu}. Thin domains, on the other hand, offer the possibility of a good compromise between computing costs and physical realism, at least in the fingering phase prior to layer formation \citep{GaraudBrummell2015}. 

\begin{figure}
\centering
\includegraphics[width=0.7\linewidth]{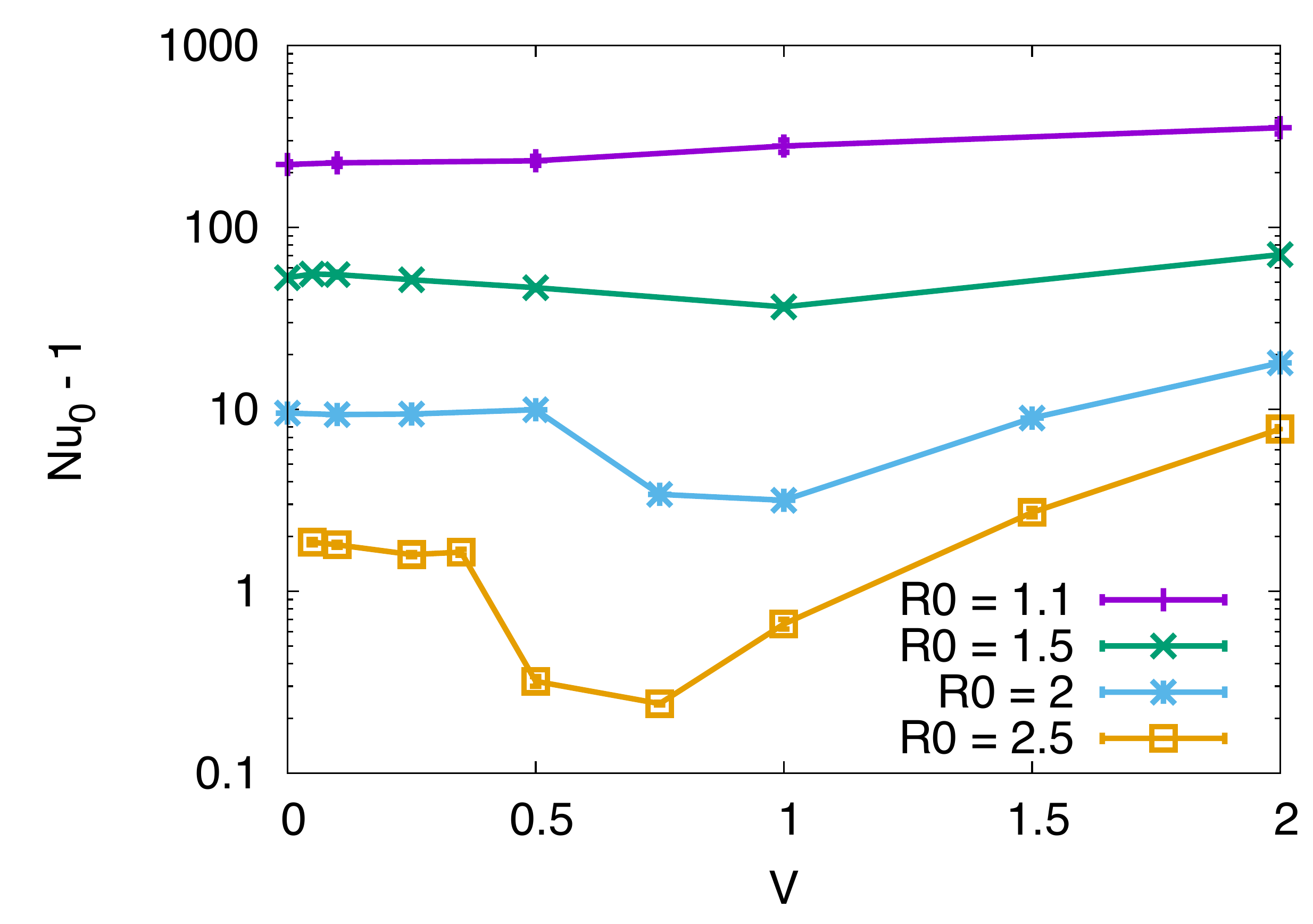}
\caption{Variation of $\Nu_0 -1 = - \langle w T \rangle_t$ with $V$ for simulations with various density ratios, $\Pran = 7$ and $\tau = 1/3$. Note that $\Nu_0$ is measured in the homogeneously turbulent phase prior to the formation of layers (if they do form). In cases where several simulations are available in different domain sizes and/or at different resolutions, we only plot the point corresponding to the largest domain and/or the highest resolution. The data for $V = 0$ is taken from \citet{Stellmach2011}. Note that the errorbars are smaller than the symbol size. \label{fig:fluxes1}}
\end{figure}

Figure \ref{fig:fluxes1} shows the average temperature Nusselt number
\begin{equation}
\Nu_0 = 1 -  \langle wT \rangle_t \, ,
\end{equation}
as a function of $V$ for fixed values of $R_0$. Note that the equivalent figure for the average particle Nusselt number $\Nu_{C,0}$ defined as
\begin{equation}
 \Nu_{C,0} =  1 - (R_0/\tau) \langle w C \rangle_t \, ,
\end{equation}
would look very similar, aside from the scaling factor of $R_0/\tau$. We see that settling has a non-trivial effect on the turbulent fluxes, sometimes enhancing them, and sometimes reducing them. The effect is particularly pronounced for high density ratio simulations, where a relatively small settling velocity can reduce the turbulent flux by a factor of a few. The latter only increases again for sufficiently large $V$. As mentioned earlier, a possible explanation for the decrease of $\Nu_0$ with $V$ at larger $R_0$ could be that the inclined sedimentary fingering modes that dominate the system dynamics at these parameters are less efficient at transporting heat than the vertically invariant standard fingering modes. What causes the increase in $\Nu_0$ for larger $V$, on the other hand, remains to be determined. Generally speaking, however, we see that the turbulent fluxes in sedimentary fingering convection are quite similar to those one would obtain in the non-sedimentary case, at least at moderate density ratios and for non-dimensional settling velocities $V$ up to 2. This reinforces the conclusions of \citet{green1987importance}.

\begin{figure}
\centering
\includegraphics[width=0.7\linewidth]{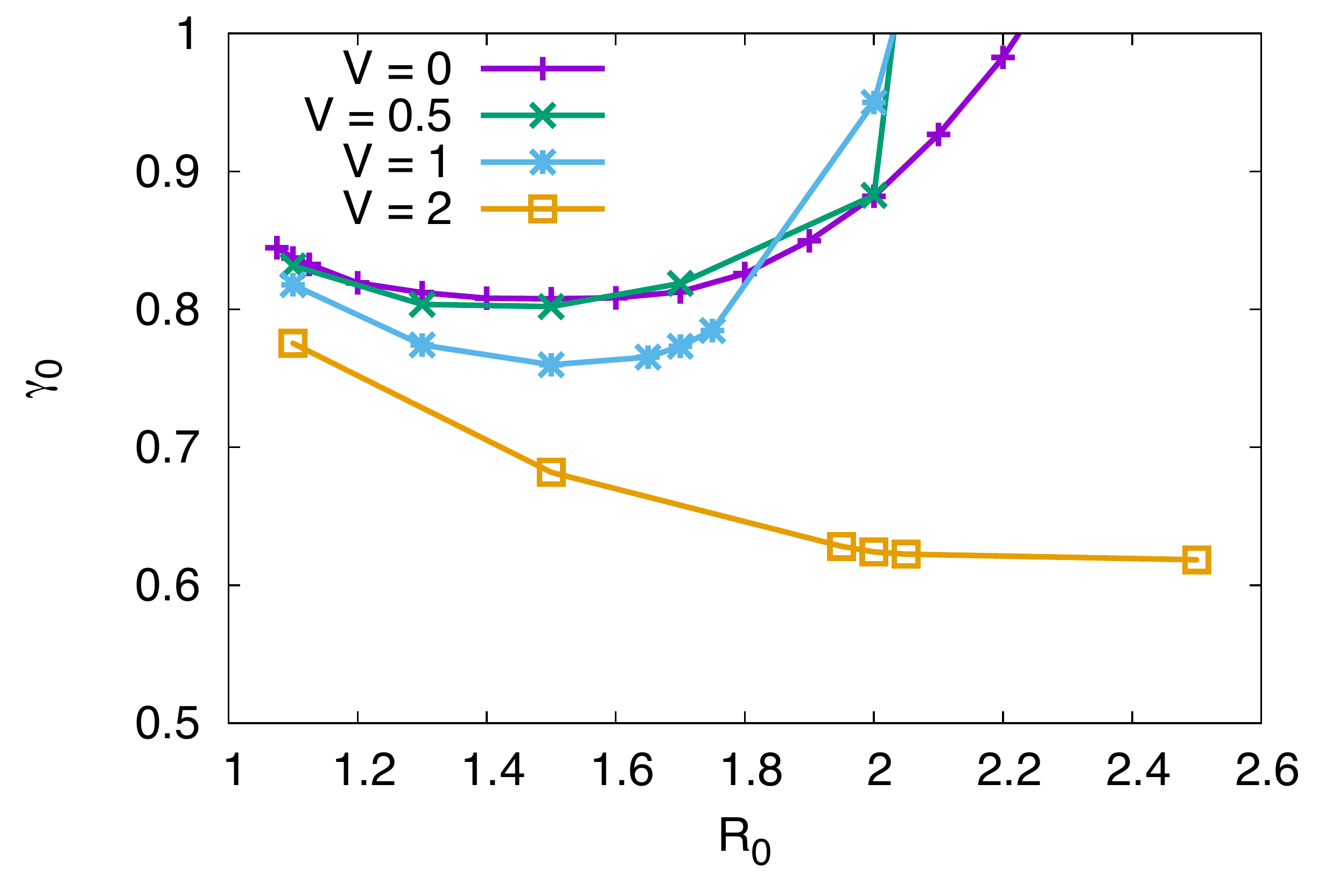}
\caption{Variation of the total flux ratio $\gamma_0$ (see equation \ref{eq:gamma}) with $R_0$ for simulations of increasing $V$. Note that $\gamma_0$ is measured in the homogeneously turbulent phase prior to the formation of layers (if they do form). In cases where several simulations are available in different domain sizes and/or at different resolutions, we only plot the point corresponding to the largest domain and/or the highest resolution. The data for $V = 0$ is taken from \citet{Stellmach2011}. \label{fig:gamma}}
\end{figure}

Figure \ref{fig:gamma} shows the average total flux ratio $\gamma_0$, which is the ratio of the average temperature flux to the average sediment flux (ignoring the contribution due to settling alone), namely 
\begin{equation}
\gamma_0 = \frac{R_0}{\tau} \frac{\Nu_0 }{ \Nu_{C,0}} \, .
\label{eq:gamma}
\end{equation} 
As discussed in Section \ref{sec:intro1}, the variation of $\gamma_0$ with $R_0$ controls the dynamics of the layer-forming $\gamma$-instability. We see that at $\Pran = 7$, $\tau = 1/3$, and $V = 0$, $\gamma$ decreases with $R_0$ for $R_0 < 1.6$. This roughly remains true for setting velocities up to $V = 1$, but changes dramatically for $V = 2$: at this larger settling rate, the range of density ratios for which $\gamma_0$ is a decreasing function of $R_0$ increases significantly.  Hence, even though the turbulent fluxes are roughly the same in the sedimentary case and the thermohaline case, the small differences can lead to a much larger change in the flux ratio. 
This has important consequences for layer formation in sedimentary fingering convection, as demonstrated below. 

\section{Layering instabilities in sedimentary fingering convection}\label{sec: Extension of Radko 2003}

\subsection{Generalization of Radko's theory}\label{sec: mft}


Having found that layers can emerge in sedimentary fingering convection in parameter regimes where they do not appear when $V = 0$, we extend the mean field theory introduced by \citet{radko2003mechanism} \citep[see also][]{Traxler2011a} to account for the effect of sedimentation, to see if we can explain our numerical findings. We begin by averaging the temperature and sediment concentration fields horizontally as
\begin{equation}
\bar T(z,t)= \frac{1}{L_xL_y} \iint T(x,y,z,t)dx dy \mbox{  and }  \bar C(z,t)=\frac{1}{L_xL_y} \iint  C(x,y,z,t) dx dy \mbox{  .} 
\end{equation}
Assuming there is no mean vertical flow, the evolution equations for $\bar T$ and $\bar C$ are
\begin{equation}\label{eq: flux grad} \frac{\p\bar T}{\p t}=-\frac{\p F_T}{\p z},  \ \ \ \  \frac{\p\bar C}{\p t}=-\frac{\p F_C}{\p z}+V\frac{\p\bar C}{\p z},
\end{equation}
where 
\begin{eqnarray}
F_T=\overline{wT}-\frac{\p\bar T_{\rm tot}}{\p z} , \\
F_C=\overline{wC}-\tau \frac{\p\bar C_{\rm tot}}{\p z} , 
\end{eqnarray}
where $\bar{T}_{\rm tot}(z,t)=z+\bar T(z,t)$ is the total horizontally-averaged temperature field (background plus perturbations) and similarly $\bar C_{\rm tot}(z,t) = z R_0^{-1} + \bar C(z,t)$. Note that the turbulent temperature and sediment fluxes  
\begin{equation}
\overline{wT} = \frac{1}{L_xL_y}  \iint wT dxdy  \mbox{   and  } \overline{wC} = \frac{1}{L_xL_y}  \iint wC dxdy
\end{equation} 
are functions of both $z$ and $t$. 

Following \citet{radko2003mechanism}, we now write $F_T$ and $F_C$ in terms of the non-dimensional temperature Nusselt number ${\Nu}$ and flux ratio $\gamma$ as:
\begin{equation}
\label{eq:Nu} 
F_T= - {\Nu} \frac{\p\bar T_{\rm tot}}{\p z} ,
\end{equation}
and
\begin{equation}\label{eq:fc}
F_C=\frac{1}{\gamma}F_T.
\end{equation}
We also assume, again as in \citet{radko2003mechanism}, that $\Nu$ and $\gamma$ are only functions of the basic parameters $\Pran$, $\tau$ and $V$ (which are constant), and of the local density ratio 
\begin{equation}
\label{eq: local R0} 
R(z,t) = \frac{1 + \frac{\p \bar T}{\p z}}{ R_0^{-1} + \frac{\p \bar C}{\p z}} \mbox{  ,}
\end{equation}
which can depend on time and vertical position.  Substituting (\ref{eq:Nu}) and (\ref{eq:fc}) into (\ref{eq: flux grad}) we have 
\begin{equation}\label{eq: T and C flux grad}\frac{\p\bar T}{\p t}=\frac{\p}{\p z}\left[\Nu\left(1+\frac{\p\bar T}{\p z}\right)\right],  \ \ \ \  \frac{\p\bar C}{\p t}=-\frac{\p}{\p z}\left[\frac{F_T}{\gamma}\right]+V\frac{\p\bar C}{\p z}.
\end{equation}
Using both the product rule and the chain rule, we can transform these equations into
\begin{eqnarray}
\label{eq: fluxT}
&& \frac{\p\bar T}{\p t}=\frac{\p \Nu}{\p R}\frac{\p R}{\p z}\left(1+\frac{\p\bar T}{\p z}\right)+\frac{\p^2\bar T}{\p z^2}\Nu(R),\\
\label{eq: fluxC}
&& \frac{\p\bar C}{\p t}=\frac{\p\gamma^{-1}}{\p R}\frac{\p R}{\p z}\Nu(R)\left(1+\frac{\p\bar T}{\p z}\right)+\gamma^{-1}(R)\frac{\p \bar T}{\p t}+V\frac{\p\bar C}{\p z}.
\end{eqnarray}
Next, we substitute (\ref{eq: local R0}) into (\ref{eq: fluxT}) and (\ref{eq: fluxC}), and linearize the result assuming that the perturbations to the linearly stratified background state are small (i.e. $\partial \bar T / \partial z \ll 1$ and $\partial \bar C / \partial z \ll R_0^{-1}$) to obtain: 
\begin{eqnarray}
\label{eq: linear equationsT}
&& \frac{\p\bar T}{\p t}=R_0\frac{\p \Nu}{\p R}\bigg|_{R_0}\left(\frac{\p^2\bar T}{\p z^2}-R_0\frac{\p^2\bar C}{\p z^2}\right)+\frac{\p^2\bar T}{\p z^2}\Nu(R_0) \, ,\\
\label{eq: linear equationsC}
&& \frac{\p\bar C}{\p t}=R_0\frac{\p \gamma^{-1}}{\p R}\bigg|_{R_0}\left(\frac{\p^2\bar T}{\p z^2}-R_0\frac{\p^2\bar C}{\p z^2}\right)\Nu(R_0)+\gamma^{-1}(R_0)\frac{\p\bar T}{\p t}+V\frac{\p\bar C}{\p z} \, .
\end{eqnarray}

Finally we assume normal modes of the form $(\bar T, \bar C)=(\hat T, \hat C)\exp(\Lambda t+ikz)$ to get
\begin{eqnarray}
\Lambda^2 +\Lambda\left[(A_{\Nu}+\Nu_0-A_\gamma \Nu_0R_0-A_{\Nu}R_0/\gamma_0)k^2 -ik V \right] \nonumber\\
-A_\gamma \Nu_0^2 R_0k^4-ik^3V(A_{\Nu}+\Nu_0)=0 \, , \label{eq: eigenvalue}
\end{eqnarray}
where for simplicity we have introduced the constants
\begin{eqnarray}
\Nu_0 = \Nu(R_0) \mbox{,} \gamma_0 = \gamma(R_0) \, , \label{eq:nugamma} \\  
A_{\Nu}=R_0\frac{\p \Nu}{\p R}\bigg|_{R_0} \mbox{ and } A_\gamma= R_0\frac{\p\gamma^{-1}}{\p R}\bigg|_{R_0} \, .
\label{eq:anuagamma}
\end{eqnarray}
By ergodicity and the homogeneous nature of the basic instability, the quantities $\Nu_0$ and $\gamma_0$ just defined can be identified with the ones discussed in Section \ref{sec: basic flux}. 
For a given set of parameters $\Pran$, $\tau$ and $V$, all four constants can therefore be constructed from the measurements presented in Tables 1 and 2. 

In what follows, we rewrite the quadratic equation (\ref{eq: eigenvalue}) for simplicity as 
\begin{equation}
\Lambda^2  + \Lambda  \left[a  k^2 -i k V\right] -b k^4-ic k^3 V=0 \, ,
\end{equation}
where 
\begin{eqnarray}
\label{eq:coeffs}
&&a=A_{\Nu}+\Nu_0 -A_\gamma \Nu_0 R_0 - R_0 A_{\Nu}\gamma_0^{-1} \, , \nonumber \\
&&b=A_\gamma \Nu_0^2 R_0 \, , \nonumber \\
&&c=A_{\Nu}+\Nu_0 \, .
\end{eqnarray}
The solutions are
\begin{equation}
\Lambda = \frac{ - (a  k - i V) \pm [(a  k -i  V)^2 + 4b k^2 + 4i c k V ]^{1/2}  }{2}  k \, .
\label{eq:quadsol}
\end{equation}
Note that one of these roots always has a negative real part, while the other can have either a positive or a negative real part. There are two well-defined asymptotic limits of this expression: for fixed $k$, the limit of $V \ll k$, and for fixed $V$, the limit of $k \ll V$. We now look at both in turn. 

\subsection{The $\gamma-$instability}\label{sec: gamma}

As expected, (\ref{eq:quadsol}) recovers the $\gamma$-instability theory of \citet{radko2003mechanism} exactly\footnote{To be precise, \citet{radko2003mechanism} defined $\gamma$ as the ratio of the turbulent fluxes, and neglected the diffusive fluxes altogether. Meanwhile, we include the diffusive fluxes in the definition of $\gamma$, but the theory is otherwise identical.} when $V = 0$, with a growth rate
\begin{equation}
\Lambda = \frac{ - a   \pm \sqrt{a^2 + 4b}  }{2}  k^2  \mbox{  .}
\label{eq:radkomode}
\end{equation}
 As discussed by \citet{radko2003mechanism}, a necessary  condition for instability in this case is $A_\gamma > 0$ (equivalently, $b>0$), which requires that $\gamma$ be a decreasing function of the density ratio.  The layer-forming instability in the context of standard fingering convection is therefore commonly known as the $\gamma$-instability. Unstable modes are sinusoidal perturbations in $\bar T$ and $\bar C$ (and therefore $\bar \rho$), that increase exponentially with time. When local inversions in the density gradient first emerge, convectively unstable layers appear separated by thin stably-stratified interfaces, and the fluid overturns into a fully-formed staircase.

Note that the $\gamma-$instability suffers, in theory, from an ultraviolet catastrophe \citep{radko2003mechanism}, since $\Lambda \propto k^2$ (see equation \ref{eq:radkomode}). In practice, however, two effects not accounted for in the basic theory alleviate the problem. On the one hand, the theory is expected to fail on length- and time-scales commensurate or smaller than those associated with the basic fingering instability, since it relies on spatio-temporal averages of the fluxes over these scales. Discrepancies between the predictions of mean-field theory and DNS for mean-field modes with large wavenumbers were already pointed out by \citet{SternSimeonov02}, and identified more clearly by \citet{Traxler2011a}, in the context of the excitation of gravity waves by the collective instability first discussed by \citet{stern1969sfa}. On the other hand, the $\gamma-$instability theory also fails on scales somewhat larger than the finger scale. Indeed, \citet{Traxler2011a} showed that the aforementioned large-scale gravity waves grow at the same time as the $\gamma$-instability, and the fastest-growing gravity-wave modes, which have a well-defined vertical scale, essentially filter out all layering modes of smaller scale. As a result, the emerging staircase has an initial layer height that is at least $100d-150d$ or greater, and can only be seen in simulations run in sufficiently tall domains (see Section \ref{sec: Comparing results to linear sim} for more on this issue).

It has long been known that in common geophysical and laboratory systems (e.g. heat and salt, salt and sugar, etc.), $\gamma(R)$ is a non-monotonic curve (see Figure \ref{fig:gamma} for instance), containing both a decreasing section for low $R$, followed by an increasing section at larger $R$ \citep{turner1967,LambertDemenkow1972}. This limits the range of density ratios over which one may expect to observe spontaneous layer formation to $R \in [1,R_{\rm crit}]$ where $R_{\rm crit}$ is the density ratio for which $\gamma(R)$ is minimum.  As shown by \citet{schmitt1979fgm}, the non-monotonicity of $\gamma(R)$ in standard fingering convection can be explained, at least qualitatively, by considering the transport induced by the linear fingering modes. To obtain more quantitative estimates of $R_{\rm crit}$, \citet{Stellmach2011} ran numerous ``salt fingering" direct numerical simulations with $\Pran=7$ and $\tau=1/3$ (and no settling), as well as with $\Pran = 7$ and $\tau = 0.01$. Their results for $\Pran = 7$ and $\tau = 1/3$ are shown in Figure \ref{fig:gamma}, and reveal that $R_{\rm crit} \simeq 1.5$ at these parameters. They also found that $R_{\rm crit} \simeq 4$ for $\tau = 0.01$. Beyond these critical density ratios, standard fingering convection is stable to the $\gamma-$instability, which means that any horizontally-invariant perturbation in the temperature or concentration fields is exponentially damped. 

For sedimentary fingering convection, but in the limit of $V \ll k$, a simple asymptotic expansion of $\Lambda$ yields 
\begin{equation}
\Lambda = \frac{-a \pm  \sqrt{ a^2 + 4 b}}{2} k^2 \pm \frac{ikV}{2} \frac{ 2c-a \pm  \sqrt{ a^2 + 4 b}      }{  \sqrt{ a^2 + 4 b} } + k^2 O(V^2/k^2 ) \mbox{  .}
\label{eq:approximatexp}
\end{equation}

This shows that, at the lowest order, the real part of $\Lambda$ is simply the growth rate of a standard $\gamma-$mode, while the first-order correction due to settling (for small $V$) merely amounts to a vertical phase propagation with velocity proportional to $V$. Corrections to the actual growth rate (i.e. the real part of $\Lambda$) only appear at order $V^2$, and are therefore negligible when $V$ is small. This finding, combined with the results of Section \ref{sec: basic flux} that show that the temperature and sediment fluxes in the limit of $V \ll 1$ are very similar to the ones obtained in the case of $V = 0$ for small-to-moderate $R$, demonstrates that the $\gamma$-instability exists in sedimentary fingering convection at low settling velocities at more-or-less the same density ratios as in the standard fingering case, and grows with more-or-less the same growth rate. The only difference lies in the slow translational motion of the layering modes, which is not present when $V = 0$. 

\begin{figure}
\centering
\includegraphics[width=\textwidth]{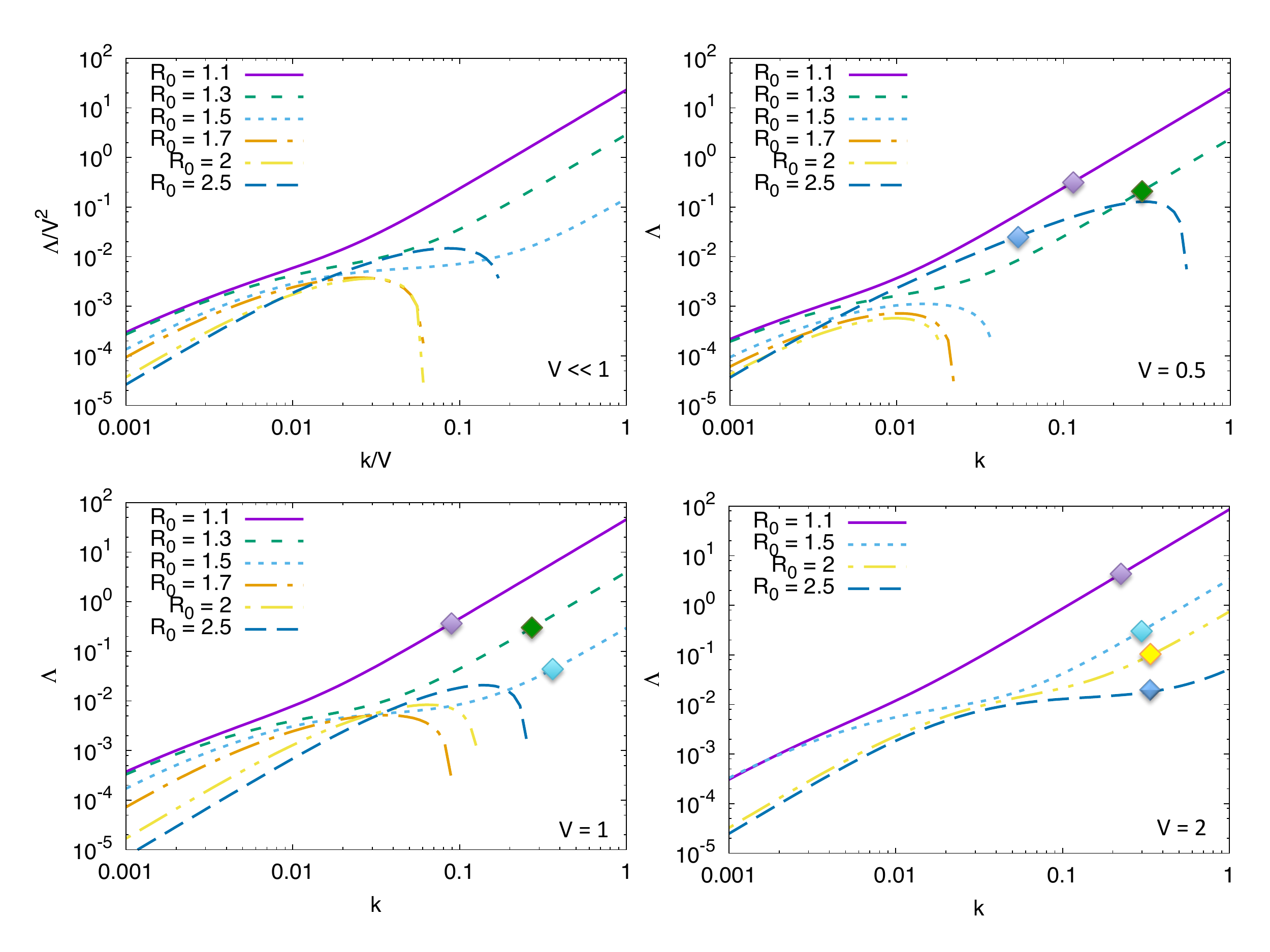}
\caption{Variation of the real part of $\Lambda$ with $k$, based on values of the parameters in (\ref{eq:nugamma})-(\ref{eq:anuagamma}) estimated from the numerical results reported in Tables 1 and 2, for various values of $V$ and $R_0$, at $\Pran = 7$ and $\tau  = 1/3$ (see main text for detail). In the top left panel, the small $V$ limit is shown in which $\Lambda$ scales with $V^2$, and $k$ with $V$. In the other panels, cases with $V = 0.5$, $V= 1$ and $V=  2$ are shown. Note how whenever $\gamma$ is a decreasing function of $R$ (in Figure \ref{fig:gamma}), $\Lambda(k)$ increases monotonically and the theoretical unstable modes suffer from an ultraviolet catastrophe reminiscent of the $\gamma-$instability. In practice, the theory fails when the mean-field mode growth rate exceeds the fastest fingering mode growth rate, or when its vertical wavenumber exceeds the fingering mode wavenumber (assuming the latter is non zero). Symbols have been added to mark the point in the curve above which the mean-field theory is no longer applicable, for $V = 0.5$, $V = 1$ and $V = 2$. Meanwhile, when $\gamma$ is an increasing function of $R$, there are also unstable modes, this time with a well-defined maximum in $\Lambda(k)$. These modes are excited by the sedimentary layering instability. \label{fig:lambdak}}
\end{figure}

For larger values of $V/k$, since the asymptotic expansion (\ref{eq:approximatexp}) is no longer valid, we evaluate $\Lambda$ exactly using (\ref{eq:quadsol}) and the data of Tables 1 and 2. The results are presented in Figure \ref{fig:lambdak}, which shows the real part of $\Lambda$ (for the root with positive real part) as a function of $k$, for various values of $V$ and $R_0$, for $\Pran = 7$ and $\tau = 1/3$. Each curve is obtained in the following way: for given values of $V$ and $R_0$, we first calculate the corresponding values of $\Nu_0$ and $\gamma_0$ using the fluxes in Tables 1 and 2, and calculate the values of $A_{\Nu}$ and $A_{\gamma}$ by finite differences on the same data\footnote{Note that the estimates of the growth rates are much more accurate for the curves $R_0 = 1.7$, $V = 1$ and $R_0 = 2$, $V = 2$, than for all the other ones. This is because for these two particular datasets we have used simulations at nearby values of the density ratio to estimate $A_\Nu$ and $A_\gamma$, see Section \ref{sec: Comparing results to linear sim} and Table 3 for detail. For all the other curves, $\Lambda$ could be off by a factor of order unity owing to the error made in approximating the derivatives $A_\gamma$ and $A_\Nu$ with finite differences using fairly separated points, although the general shape of the curves is expected to be correct.} . We then compute $\Lambda$ using (\ref{eq:quadsol}) for each value of $k$. The curves for the $V \ll 1$ panel are obtained using the data for $V = 0.1$.

Comparing Figures  \ref{fig:gamma} and \ref{fig:lambdak}, we see that whenever $\gamma_0$ is a decreasing function of $R_0$ (in Figure \ref{fig:gamma}), there is an unstable mode in Figure \ref{fig:lambdak} whose growth rate $\Lambda$ increases monotonically with $k$ and suffers from the same ultraviolet catastrophe as the original $\gamma$-modes. This shows that the $\gamma-$instability persists and operates more or less the same way in the sedimentary case as in the non-sedimentary case. An important consequence of this result is that since $\gamma$ appears to be monotonically decreasing at larger settling velocities (e.g. for $V = 2$), at least for the range of density ratios tested, we predict that layer formation should be relatively ubiquitous in that limit.

\subsection{The sedimentary layering instability}\label{sec: sedlayer}

In Section \ref{sec: basic flux} we found that layering occurred for a simulation with $R_0 = 1.7$ and $V = 1$, when it did not for $V = 0$. Inspection of Figure \ref{fig:gamma} shows that $\gamma_0$ is an increasing function of $R_0$ at these parameters, so the layers that form cannot be due to the $\gamma$-instability. This suggests that another layering mechanism must be at play for higher density ratios and $V \neq 0$. In order to study it, we first note that by rescaling $\Lambda=V^2\hat \Lambda$ and $k=V\hat k,$ equation (\ref{eq: eigenvalue}) becomes
\begin{equation}
\hat\Lambda^2  +\hat\Lambda  \left[a\hat k^2 -i\hat k\right] -b\hat k^4-ic\hat k^3=0 \, . \label{eq: scaled eigenvalue}
\end{equation}
This rescaling is only possible when $V \neq 0$. We now see that 
all explicit dependence on $V$ has disappeared from equation (\ref{eq: scaled eigenvalue}), although the solution $\hat \Lambda$ can still depend implicitly on $V$ through the constants $\Nu_0$, $\gamma_0$, $A_\gamma$ and $A_{\Nu}$ (since the turbulent fluxes from which they are constructed depend on the settling velocity). 

Assuming $\hat k \ll 1$, or equivalently $k \ll V$, and letting $\hat \Lambda = \alpha_1 \hat k+ \alpha_2 \hat k^2 + O(\hat k^3)$ into (\ref{eq: scaled eigenvalue}) allows us to solve for the complex coefficients $\alpha_j$ to get an approximation of $\hat \Lambda$ for small $\hat k$. The resulting asymptotic approximations for the two solutions $\hat \Lambda_1$ and $\hat \Lambda_2$ are 
\begin{eqnarray}
\label{eq: ass exp}
\hat \Lambda_1 =-c\hat k^2 + O(\hat k^3) \, , \nonumber \\ 
\hat \Lambda_2= i \hat k + (c-a)\hat k^2 + O(\hat k^3) \, , 
\end{eqnarray} 
where $a$, $b$, and $c$ were defined in (\ref{eq:coeffs}). Note that (\ref{eq: ass exp}) is equivalent to 
\begin{eqnarray}
 \Lambda_1  =-c k^2   + O( k^3/V) \, , \nonumber \\ 
\Lambda_2 = i  kV  + (c-a) k^2  + O(k^3/V) \, ,
\end{eqnarray} 
and could also have been obtained by a direct asymptotic expansion of (\ref{eq:quadsol}) in the small parameter $k/V$. 
We therefore see that, should unstable modes exist at low $\hat k$, their growth rate must be proportional to $\hat k^2$.

Examining (\ref{eq: ass exp}) in more detail shows that simple sufficient conditions for instability are either  
\begin{eqnarray}
c < 0 \Leftrightarrow \left. \frac{\p}{\p R}(R\Nu )\right|_{R_0} < 0 \, ,  \mbox{   or  }\\
c-a > 0 \Leftrightarrow  \left. \frac{\p}{\p R}\left(\frac{ \Nu }{\gamma} \right)\right|_{R_0} > 0 \, .
\end{eqnarray}
Hence, we predict that layers can form spontaneously in sedimentary fingering convection either if $R_0 \Nu_0$ decreases with $R_0$, or if $\Nu_0/\gamma_0$ increases with $R_0$. Note that this criterion does not explicitly depend of the value of $V$, although there is an implicit dependence on the settling velocity through its effects on the basic fluxes that are used to compute $c$ and $a$.

\begin{figure}
\includegraphics[width=\textwidth]{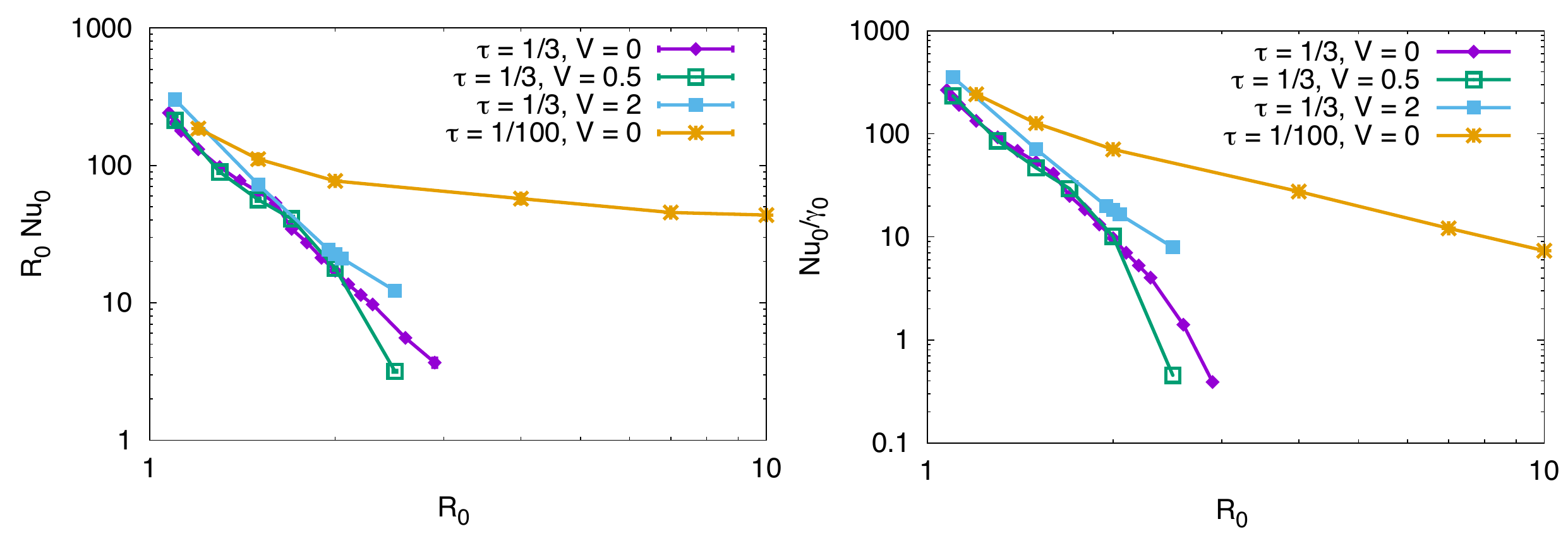}  
\caption{Variation of $R_0\Nu_0$ and $\Nu_0/\gamma_0$ with $R_0$, for $\Pran = 7$, $\tau = 1/3$ and $\tau = 1/100$. Note that $\gamma_0$ and $\Nu_0$ are measured in the homogeneously turbulent phase prior to the formation of layers (if they do form).The data for $V = 0$ is taken from \citet{Stellmach2011}. \label{fig2}}
\end{figure}

Using the flux data from non-sedimentary fingering convection of \citet{Stellmach2011}, for $\tau=1/3$ and $\Pran=7$, and for $\tau = 0.01$ and $\Pran = 7$ -- which is a good approximation to the actual flux data for small enough $V$ -- as well as our data from Section \ref{sec: basic flux} for $V = 0.5$ and $V = 2$, we can determine the signs of $c$ and $c-a$ by plotting $R_0 \Nu_0$ and $\Nu_0 / \gamma_0$ against $R_0$. Figure \ref{fig2} shows that $R_0 \Nu_0$ is indeed a decreasing function of $R_0$ for the data available over the entire range of $R_0$ (so $c < 0$), while $\Nu_0/\gamma_0$ is always a decreasing function of $R_0$ for the same data (so $c-a < 0$), for $V = 0$, $V = 0 .5$ and $V = 2$. So, the mode with growth rate $\hat \Lambda_1$ is unstable while the one with growth rate $\hat \Lambda_2$ is stable. We therefore expect a layering instability to take place for the {\it entire} fingering range for the case of $\tau = 1/3$ and for a significant part of the fingering range\footnote{Measurements of $\Nu_0$ and $\gamma_0$ from direct numerical simulations are only available for $R_0$ up to 10, while the critical density ratio for stability is $R_0 = 100$ when $\tau = 0.01$.} when $\tau = 0.01$. 

Going back to Figure \ref{fig:lambdak}, we now see that a layering instability is indeed theoretically present at all density ratios! This result is remarkable given that, by contrast, the $\gamma-$instability is limited to the region of parameter space where $\gamma$ is a decreasing function of $R$. We also see that the new sedimentary layering instability does not suffer from an ultraviolet catastrophe, and has instead a well-defined fastest-growing mode and a well-defined high-wavenumber cutoff. We now give a physical explanation of the sedimentary layering instability, and clarify why it is theoretically easier to trigger than the $\gamma$-instability. 



\subsection{The physical mechanism for the sedimentary layering instability}\label{sec: Physical Mech} 

A better understanding of the physical mechanism behind the sedimentary layering instability can be gained by looking, as in \citet{radko2003mechanism}, at the structure of the linearly unstable layering modes. For the sake of illustration, we continue to use the parameters $\Pran = 7$ and $\tau = 1/3$. Since the mode amplitude (for one of the variables) is arbitrary, we take  
\begin{equation}
\bar T(z,t)= \Re( \hat T e^{i k z+ \Lambda t}) = 0.1 \sin(kz) e^{\Lambda t}\mbox{  ,}
\end{equation}
that is, with $\hat T=- 0.1 i$. We also pick $k = 0.01$, a value for which unstable modes exist both for the $\gamma-$instability at low density ratio, and for the sedimentary layering instability for high density ratio with $V = 1$ (see Figure \ref{fig:lambdak}). 
We can then calculate $\bar C$ via (\ref{eq: linear equationsC}), which yields
\begin{equation}
 \bar C(z,t)= \Re \left[  \frac{ \Lambda +  k^2 (\Nu_0 + A_\Nu )  }{k^2 A_\Nu R_0}   \hat  T  e^{i  k z} \right]e^{\Lambda t}  \mbox{  .}
 \end{equation}

Using these $\bar T$ and $\bar C$ profiles, we then compute $R(z,t),\gamma(z,t),F_T(z,t),$ and $F_C(z,t)$ to first order in the perturbations as 
\begin{eqnarray}
&& R - R_0 = R_0\frac{\partial \bar T}{\partial z}-R_0^2\frac{\partial \bar C}{\partial z} \, , \\ 
&& \gamma-\gamma_0 = -\gamma_0^2A_\gamma \frac{R-R_0}{R_0} \, , \\
&& F_T+\Nu_0  
=-\Nu_0 \frac{\partial \bar T}{\partial z} - A_{\Nu} \frac{R-R_0}{R_0} \, , \\ 
&& F_C+\Nu_0 \gamma_0^{-1}= -\Nu_0 R_0 \gamma_0^{-1}   \frac{\partial \bar C}{\partial z} - (A_{\Nu} \gamma_0^{-1} + \Nu_0 A_\gamma + \Nu_0 \gamma_0^{-1}) \frac{R-R_0}{R_0} \, .
\end{eqnarray}
Note how, in the expressions for $F_T$ and $F_C$, we have separated out the background fluxes in the homogeneous case (on the left-hand-side), the contributions to the fluxes from variations that come from the gradients of $\bar T$ and $\bar C$ (term 1 on the right-hand-side) which are purely downgradient, and finally, terms that arise from the variation of the density ratio itself that are not necessarily downgradient (term 2 on the right-hand-side, proportional to $R-R_0$). The downgradient fluxes always have a stabilizing effect, so that the instability, if present, must arise from the behavior of the non-downgradient terms. 

\begin{figure}
\centering
\includegraphics[width=\textwidth]{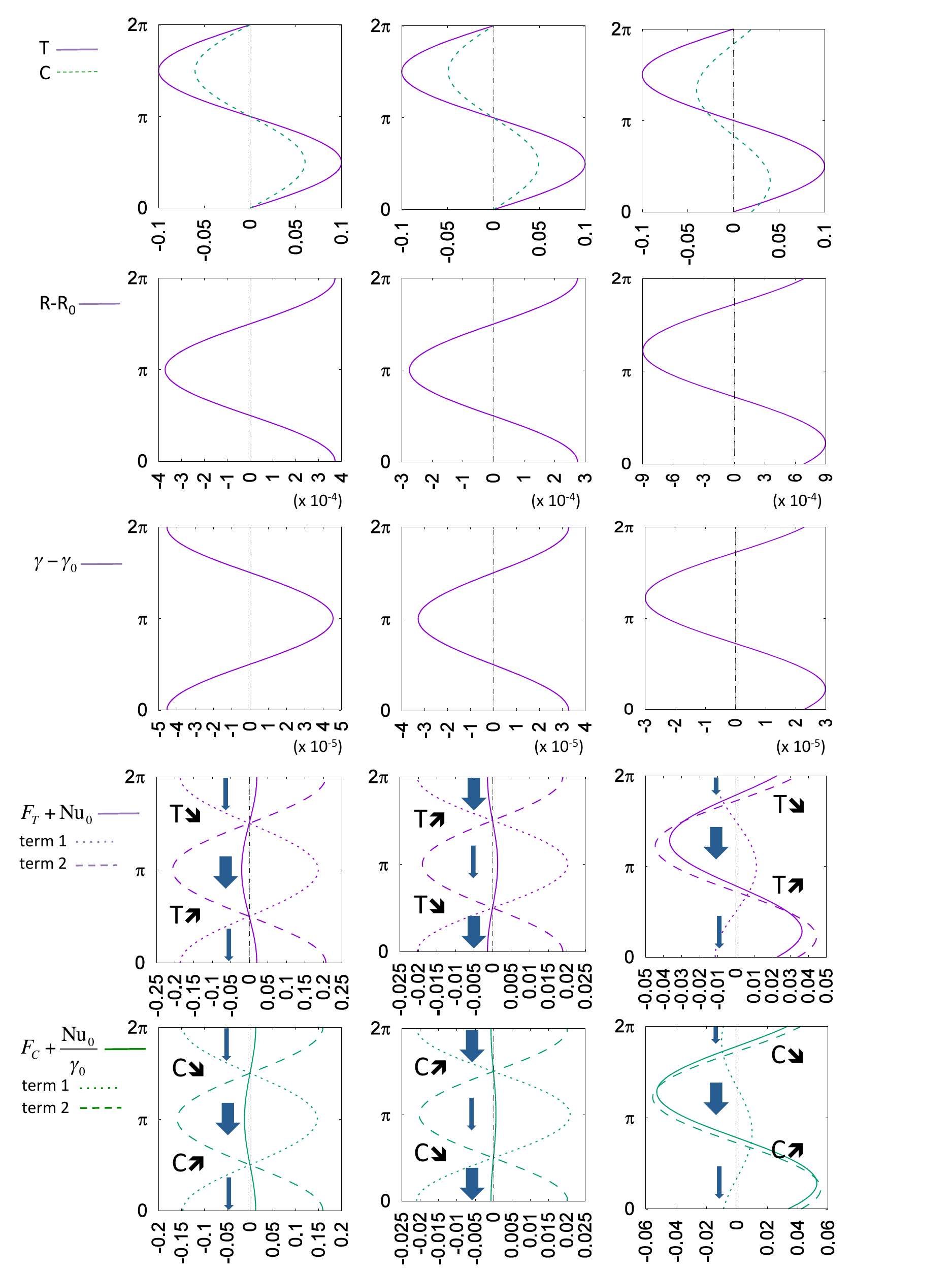}
\caption{ Illustration of the physical mechanism of the various layering instabilities, adapted from Figure 4 of \citet{radko2003mechanism}. The vertical axis in all figures shows $kz$, with $k = 0.01$. In the bottom two rows, the thickness of the arrows represents the amplitude of the total temperature and composition fluxes, to emphasize the regions of flux convergence (where the relevant quantity increases with time) and divergence (where the relevant quantity decreases with time). Left: Illustration of the $\gamma-$instability, with $R_0=1.1$, $V=0$; Center: No instability for density ratio $R_0=1.7$, $V=0$; Right: The new layering instability for sedimentary fingering convection, with density ratio $R_0=1.7$ and $V=1$.   \label{fig4}}
\end{figure}

Figure \ref{fig4} uses these results to illustrate the positive feedback loops driving the $\gamma$-instability and the sedimentary layering instability. This figure is inspired from the schematic in Figure 4 of \citet{radko2003mechanism}. The first two columns (from left to right) illustrate cases without settling ($V = 0$), one that is $\gamma$-unstable (left column, for $R_0 = 1.1$), and one that is not (middle column, for $R_0 = 1.7$). The last column shows a case that is $\gamma$-stable, but unstable to the sedimentary layering instability ($V = 1$, $R_0 = 1.7$).
 
In each column we see, from top to bottom, the temperature and salinity/sediment perturbations, the density ratio perturbation, the perturbation in $\gamma$, and finally, perturbations in the total temperature and composition fluxes. In the first column, which depicts a situation unstable to the $\gamma$-instability, we see that the respective variation of $R$, and hence $\gamma$, leads to net temperature and composition fluxes that reinforce the original perturbation and drive their growth. Inspecting the respective contributions of the downgradient flux (term 1, stabilizing) and the non-downgradient flux (term 2, destabilizing), we see that the two nearly cancel out, but the latter is slightly larger because $A_\gamma > 0$, hence driving the instability. In the second column, which is $\gamma$-stable (hence $A_\gamma < 0$), we also see that the term 1 and term 2 contributions to the fluxes nearly cancel out but this time it is the downgradient contribution that is slightly larger. As a result, the instability is quenched. 

Finally, when $V>0$, the temperature and sediment concentration perturbations are out of phase (see top right panel), the latter being shifted downward slightly because of settling. The density ratio and flux ratio perturbation profiles look superficially quite similar to the ones in the $\gamma-$stable case (which is not surprising, since all parameters except $V$ are the same in the last two columns). However, looking at the downgradient (term 1) and non-downgradient (term 2) contributions  to the total fluxes, we now see that the phase shift between the temperature and sediment field also causes a phase shift between two flux contributions, and the two no longer nearly cancel out. Instead, the non-downgradient contribution dominates, and creates a different positive feedback loop that amplifies the perturbations. Also note how the regions of flux convergence and divergence are slightly offset from the maxima and minima of the original temperature and sediment profiles, which explains how the feedback loop also causes the perturbations to travel vertically slowly. 
 
\section{Comparison of the results with direct numerical simulations}\label{sec: Comparing results to linear sim}

We now verify our theoretical findings on the two kinds of layering instabilities by comparison with 3D DNS, and study the evolution of the layers after they form. As in Section \ref{sec: basic flux}, we solve equations (\ref{eq:non_dimu}) to (\ref{eq:non_dimm}) in triply-periodic domain with $\Pran=7$ and $\tau=1/3$. We first focus on two specific cases, then summarize our general findings for all parameter values.

\subsection{Case with $R_0 = 2$, $V = 2$}
\label{sec:case1}

The first case investigated has $R_0=2$ and $V = 2$, and is run in a tall domain of size $100\times100\times200$ (see Table 2 for detail). It is in a regime that is unstable to the $\gamma$-instability since, at these parameter values, $\gamma$ is a decreasing function of the density ratio (see Figure \ref{fig:gamma}). 

\begin{figure}
\centering
\includegraphics[width=0.5\linewidth]{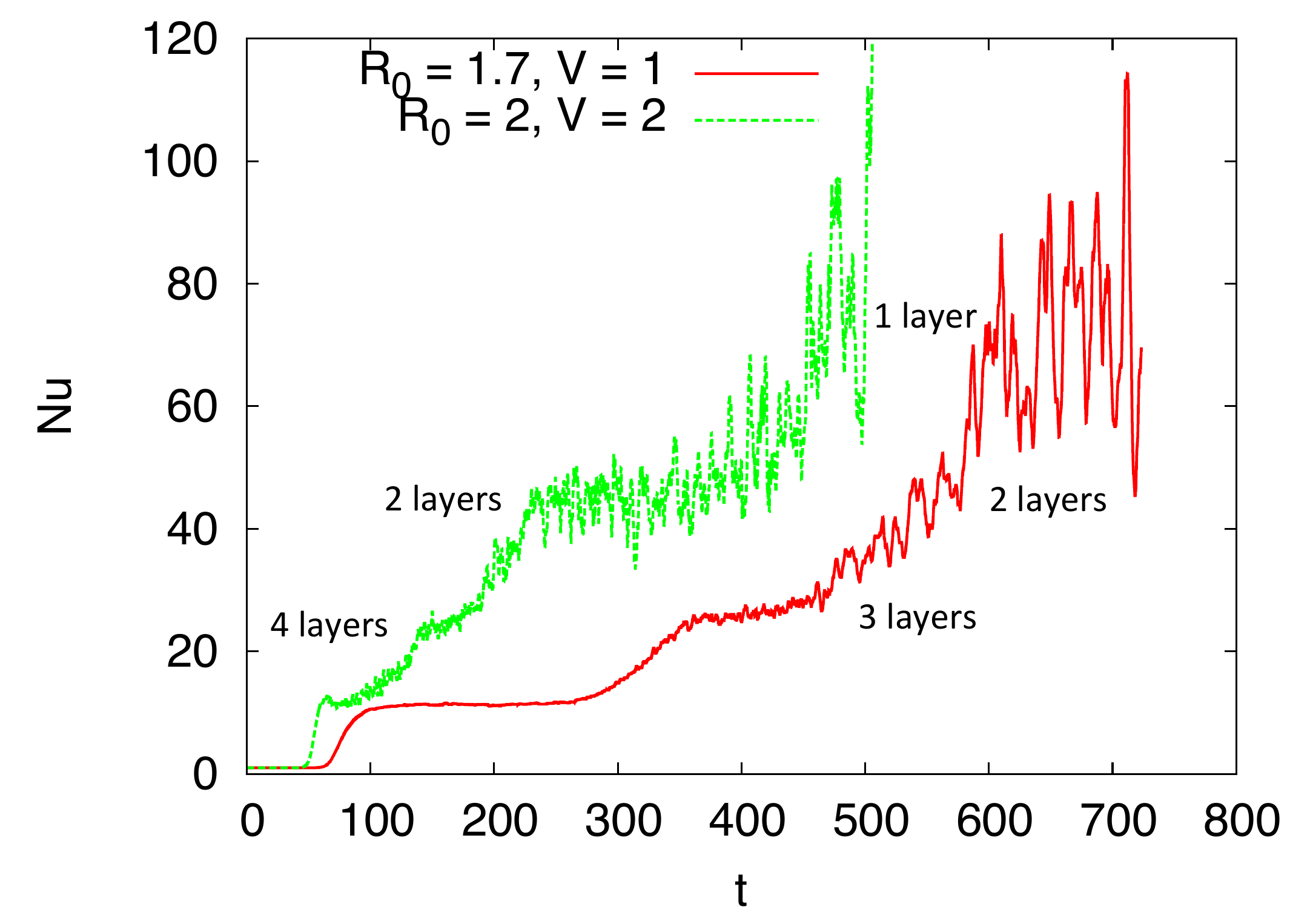}
\caption{Nusselt number $\Nu(t) = 1 - \langle wT \rangle$ as a function of time for the large-domain simulations with $R_0 =  2$ and $V = 2$ (see Section \ref{sec:case1}) and $R_0 =  1.7$ and $V = 1$ (see Section \ref{sec:case2}). In both cases, $\Pr = 7$ and $\tau  =1/3$. Each layer merger is accompanied by a step-wise increase in the Nusselt number.  The number of layers after each merger is indicated for each curve. \label{fig:Nusselt}}
\end{figure}

\begin{figure}
\centering
\includegraphics[width=\linewidth]{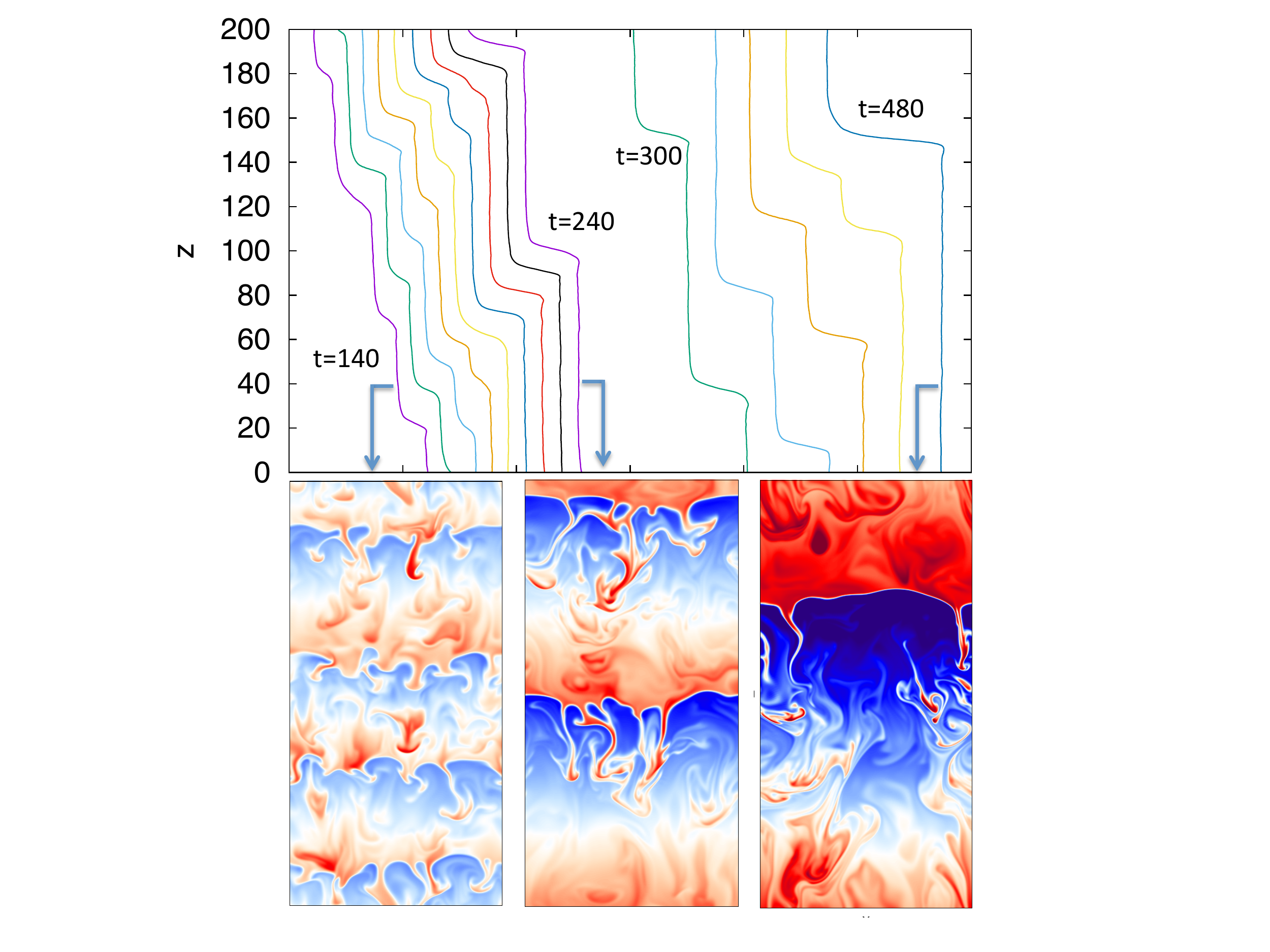}
\caption{Top: Horizontally averaged density profiles $\bar \rho_{\rm tot}(z) = (R_0^{-1} - 1)z - \bar T(z)+ \bar C(z)$ in the tall-domain simulation with $R_0 = 2$, $V = 2$, $\Pr = 7$ and $\tau = 1/3$, taken at regular time intervals between $t = 140$ and $t = 240$, then again from $t = 300$ to $t = 480$. The profiles are staggered horizontally for clarity. At early times, we see the four well-defined layers, which eventually merge into two. After some time, the layers further merge leaving only one layer. Note how the interfaces move upward as a function of time. Bottom: snapshots of the sediment concentration field at the times shown by the arrows. \label{fig:mergers}}
\end{figure}

Figure \ref{fig:Nusselt} shows the instantaneous temperature Nusselt number, $\Nu(t) = 1 - \langle wT \rangle$. It is consistent, at very early times, with the exponential growth of the fastest-growing sedimentary fingering mode, then saturates briefly, then grows again as a result of layer formation. Figure \ref{fig:mergers} shows a snapshot of the sediment concentration in this simulation at time $t = 140$, together with a density profile taken at the same time. We clearly see in both cases four well defined layers, suggesting an initial layer height around $50$. Note that the initial layer height cannot be predicted by theory at these parameter values. Indeed, there is no fastest-growing layering mode as the $\gamma$-instability suffers from an ultraviolet catastrophe (see Figure \ref{fig:lambdak}).

The four layers merge down to two then finally one layer between $t = 140$ and $t = 480$, as seen in Figure \ref{fig:mergers}. With each layer merger, the Nusselt number (as well as the total kinetic energy) increases (see Figure \ref{fig:Nusselt}). The rapid mergers are unexpected, since they were not observed in the simulations of non-sedimentary fingering convection by \citet{radko2003mechanism} or \citet{Stellmach2011}. This may be a property associated with sedimentation only. Indeed, we see that the interfaces migrate upward with time, and the mergers appear to be due to the convergence of two interfaces moving at different speeds (see Figure \ref{fig:mergers}). 

\begin{figure}
\centering
\includegraphics[width=.9\linewidth]{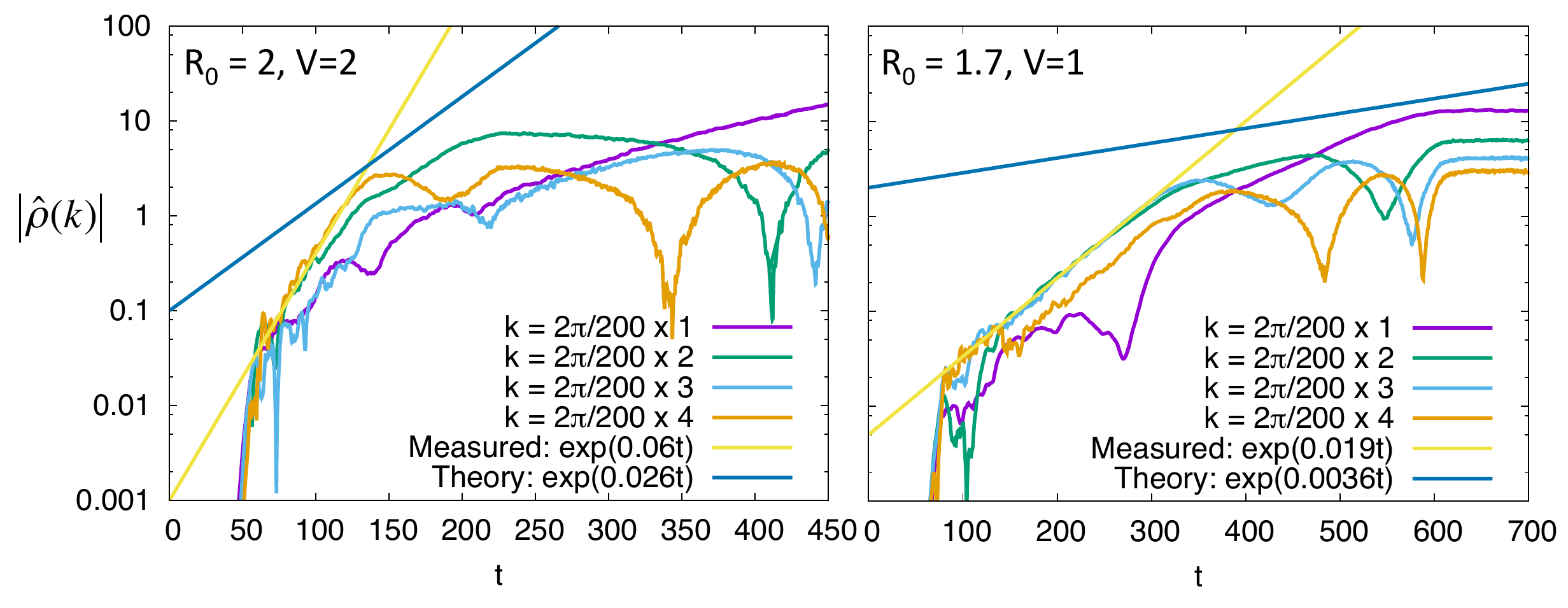}
\caption{Amplitude of the first four Fourier modes of the horizontally-averaged density perturbation profiles, as a function of time (see main text for detail). On the left is the data for the case with $R_0 = 2$, and $V = 2$ (see Section \ref{sec:case1}). On the right is the case with $R_0 = 1.7$, and $V = 1$ (see Section \ref{sec:case2}).  Also shown in both cases are the fitted growth rates of the initially dominant layering mode, as well as the theoretical growth rate of the same mode.}
 \label{fig:layeramps}
\end{figure}

To analyze the results of the simulation more quantitatively, we perform a Fourier transform of the horizontally-averaged density perturbations, as in 
\begin{equation}
\bar \rho(z,t) = - \bar T(z,t)  + \bar C(z,t) = \sum_{k = -N_k}^{N_k} \hat \rho_k \exp(ikz)\, ,
\end{equation}
and plot, as a function of time, the amplitude $|\hat \rho_k|$ of the first 4 modes (here, $k_1 = 2\pi/200$, $k_2 = 4 \pi/200$, $k_3  = 6\pi/200$ and $k_4 = 8\pi/200$). The results are shown in Figure \ref{fig:layeramps}. We see that the mode that initially dominates has $k = k_4$, which is not surprising since we have seen that the staircase initially forms with 4 steps. The measured growth rate of this layering mode, obtained by fitting this data, is $\Lambda_{\rm obs} \simeq 0.06$. 

In order to generate a reasonably accurate prediction for the theoretical growth rate of the layering modes, we need accurate estimates of $A_{\Nu}$ and $A_\gamma$. To obtain them, we run two additional simulations for $R_0 = 1.95$ and $R_0 = 2.05$, with $V = 2$, measure the turbulent fluxes (see Table 3), and construct $A_\Nu$ and $A_\gamma$ at $R_0 = 2$ using centered finite differences (see equation \ref{eq:anuagamma}). We then evaluate the theoretical growth rate of the $4$-layer mode using equation (\ref{eq:quadsol})  for $k = k_4$, and find that $\Lambda_{\rm theor} = 0.026$, which is a little bit under half of the measured growth rate. The discrepancy is rather surprising given that, to date, the $\gamma-$instability theory has been very successful in explaining the layering mode growth rates in a variety of other studies \citep[e.g.][]{Stellmach2011,rosenblumal2011}. This problem is discussed in more detail in Section \ref{sec:caseall}.
\begin{table}
\begin{center}
\begin{tabular}{ccccc}
H:  & $V = 1 $ &   &   &          \\
$R_0$  & $-\langle wT \rangle_t $ & $-\langle wC \rangle_t $  & $L_x\times L_y \times L_z$ &  $N_x \times N_y \times N_z$   \\
1.65 &13.10$\pm$	0.18 &18.22$\pm$0.22 & $100\times100\times200$ &	$192\times 192\times384$  \\
1.75  & 	7.96$\pm$0.08	& 11.23$\pm$0.10& $100\times100\times200$ &	$192\times 192\times384$ \\
\\
I:  & $V = 2 $ &   &   &         \\
$R_0$  & $-\langle wT \rangle_t $  & $-\langle wC \rangle_t $ & $L_x\times L_y \times L_z$ &  $N_x \times N_y \times N_z$  \\
1.95	&11.49$\pm$	0.37	& 19.71$\pm$ 0.47 &  $100\times100\times200$ &	$192\times 192\times384$ 	\\
2.05	&9.28$\pm$	0.26 &	16.35$\pm$	0.29	&  $100\times100\times200$ &	$192\times 192\times384$ 	\\
\end{tabular}
\end{center}
\caption{As in Tables 1 and 2. } 
\end{table}

\subsection{Case with $R_0 = 1.7$, $V = 1$}
\label{sec:case2}

The second case investigated is in a regime that is stable to the $\gamma$-instability but unstable to the sedimentary fingering instability discussed in Section \ref{sec: sedlayer}. The parameters selected are $R_0 =1.7$ and $V = 1$, and the simulation is run, this time, in a domain of size $200 \times 100 \times 200$ (see Table 2). Note that this simulation is ever-so-slightly under-resolved once in the 2- and 1-layered phases, but this has little impact on the result (see for instance the comparison of the low- and high- resolution runs in the similar simulations at $R_0 = 2$ and $V = 2$).

The staircase forms with 3 layers, that can be seen in snapshots of the sediment field (not shown). The $k = k_3$ mode is also clearly visible as one of the two dominant layering modes in Figure \ref{fig:layeramps}, and grows with rate $\Lambda_{\rm obs} = 0.019$. To compute the theoretical growth rate of this mode, we use, as in the previous section, additional small-domain runs at $R_0 = 1.65$ and $R_0 = 1.75$ to get accurate estimates of $A_\gamma$ and $A_{\rm Nu}$ (see Table 3). We find that this mode is just beyond the edge of the linearly unstable regime according to our theory, which again shows some discrepancy between theoretical predictions and simulations. We do see, however, that the $k = k_2$ mode also grows nearly at the same rate and with the same amplitude as the $k = k_3$ mode, suggesting that the two may in fact be interacting with one another\footnote{The same kind of interaction is present in other cases unstable to the sedimentary layering instability, notably the run with $R_0 = 1.5$, $V = 0.5$ in the tall, thin domain.}. The 2-layer mode is unstable according to our theory, with a predicted growth rate of $\Lambda_{\rm theor} = 0.0036$, which is significantly smaller than the observed one. 

Once formed, the layers behave exactly as they do in the previous case with $R_0  = 2$ and $V = 2$: the interfaces move slowly upward, and catch up with one another thus causing mergers. Each merger is associated with an increase in the turbulent fluxes, as seen in Figure \ref{fig:Nusselt}. The ultimate state of the system in this case is also one with a single layer. 

\subsection{All cases}
\label{sec:caseall}

Our general expectations, based on the theories developed, is that layers should always form provided simulations are run in sufficiently large domains and for a sufficiently long time. In cases that are $\gamma$-unstable, the layers should have small initial separations and grow rapidly (hence should always be found), while in cases that are $\gamma$-stable yet unstable to the sedimentary layering instability, initial layers should be fairly tall and grow comparatively slowly (hence may not appear in small-domain simulations that are run for a relatively short time only). Tables 1 and 2 summarize the outcome of all the runs we have performed, list whether layers are seen to form, and if so, how tall they are at first, at what rate the dominant layering mode appears to grow, and finally, what is the theoretical growth rate of that mode (for the 3D simulations only). We can analyze the general trends that emerge from the data in the light of our theoretical expectations. 

Since $R_0 = 1.1$ is always $\gamma-$unstable, it is perhaps surprising at first to see that very few of the simulations exhibit layer formation at this density ratio, the only exceptions being the tall domain ones. This, however, is reasonably well understood. As discussed by \citet{Stellmach2011}, for standard fingering convection at $R_0 = 1.1$, a collective instability of the fingers concurrently drives gravity waves on a scale of about 100$d$-150$d$, which filter out all layers on that scale or smaller. As a result, only layers with sufficiently low $k$ can grow, and this can only happen in a sufficiently large domain.  Unfortunately, running larger-domain simulations is extremely costly, because of the large Reynolds number of fingering convection at this low density ratio. The same is true for $V > 0$, which is why we do not see layers form in any of our simulations at $L_z =  100$. 

A similar logic applies to the runs at $R_0 = 1.3$ and $R_0 = 1.5$. In this case, however, because the density ratio is a little larger, we were able to obtain results in 2D and/or in tall, thin domains of size $L_z \ge 400$. As predicted, we find layer formation in sufficiently tall domains even when they do not form in shorter ones at the same parameters (see, for instance, the various runs at $R_0 = 1.3$ and $R_0 = 1.5$, with $V = 0.5$). Interestingly, we also observe the formation of layers for large enough settling velocity ($V = 1)$ in the smaller domains. We believe that this could be due to two concurrent effects. On the one hand, we see from Figure \ref{fig:gamma} that the slope of $\gamma_0(R_0)$ at $R_0 = 1.3$ becomes progressively steeper as $V$ increases (see Figure \ref{fig:gamma}), so $A_\gamma$ increases. This in turn implies that layers grow a little more rapidly for larger $V$ (see also Figure \ref{fig:lambdak}). On the other hand, we also see from Table 1 that the Nusselt number is somewhat smaller at $V=1$ than at $V = 0$ at $R_0 = 1.3$ and $R_0 = 1.5$. Since the growth rate of the collective instability decreases as $\Nu_0$ decreases, the wave-field is also expected to be weaker -- this is indeed verified by inspection of the simulations. Combining the larger growth rate of the layering modes with the weakened gravity-wave filter, we may therefore indeed expect to see layers emerge in small domains from the $\gamma-$instability for larger $V$, when they do not at smaller $V$. 

At density ratios of $R_0 = 1.7$ and larger, the $\gamma$-instability is inactive (except at the largest settling velocity, $V = 2$). Focussing on the cases with $V \le 1$ for now, we see the following trends. Layers are not observed in small $V$ runs. Then, beyond a critical value of $V$, layers appear, but only in sufficiently-tall domains as their initial separation is large. Finally, as $V$ continues to increase, the initial layer height decreases and the layer growth rate increases. These results are qualitatively consistent with the growth rate predictions of Figure \ref{fig:lambdak} for cases which are not $\gamma$-unstable. Recall that, by the asymptotic solution, the sedimentary layering instability has a high-wavenumber cutoff, which is  proportional to $V$ for small $V$. The fastest-growing mode growth rate is proportional to $V^2$. Hence, as $V$ increases, layers can fit in smaller domains and grow more rapidly, which is what we see. 

In each case where a staircase was seen to form, we have measured the growth rate of the dominant layering mode, and (except for the 2D runs) have calculated its corresponding theoretical growth rate using the theory developed in Section \ref{sec: Extension of Radko 2003}. Looking at Tables 1 and 2, we find that the theory never fits the data particularly well -- usually overestimating the observed growth rate somewhat at low density ratios, and  underestimating it, sometimes quite significantly, for larger density ratios. We believe that these discrepancies could be due to two factors. On the one hand the growth rates of $\gamma$-unstable modes are highly sensitive to the value of $A_\gamma$, which cannot be measured very accurately within the scope of our computing allocation. On the other hand, at large density ratios the fingers are very elongated in the vertical direction, while the mean-field theory used here assumes that the fingers behave in a local manner. It is quite possible that the vertical coherence of high-density ratio fingers further participates in the positive feedback loop between the growing layering modes and the vertical fluxes, in such a way as to systematically increase the layer growth rate. This could also explain the apparent interaction between the growing layering modes, observed in Section \ref{sec:case2}. Exactly how one could model this effect theoretically remains to be determined. 

Finally, given the high computing costs of most simulations, it is very tempting to resort to 2D simulations which are much cheaper. We have already compared the basic turbulent fluxes measured in 2D and 3D simulations prior to the formation of layers and at the same parameters in Section \ref{sec:basicfluxes}, finding that the former underestimate the latter significantly. This error then propagates to the layer-formation phase, as seen in Tables 1 and 2. For instance, layers in 2D form much more quickly, and are much thinner than in 3D in the $R_0 = 1.7$ and $V = 0.5$ case. This is probably because, at these parameter values, the 2D run is already $\gamma-$unstable while the 3D run is not. Hence, one should be very careful in interpreting the outcome of 2D runs -- they provide good qualitative insight into the problem, but cannot be relied upon for quantitative work.  



\section{Application to profiles with non-constant gradients}\label{sec: Gaussian}

So far, we have focused on studying sedimentary fingering convection and the subsequent formation of layers in the presence of constant background gradients of temperature and sediment concentration. As discussed in Section \ref{sec: The Model}, this setup is mathematically rigorous and certainly convenient, but perhaps a little contrived. One may therefore wonder whether layers could also form in more natural conditions. To study this, we now look at a situation where the background temperature gradient is held constant as previously, but where the sediment concentration profile can evolve freely from a given set of initial conditions. By contrast with the previously studied setup, which allows the system to reach a turbulent statistically stationary state, this time we perform run-down numerical experiments. 

From here onward, there is no background sediment profile and $C$ denotes the {\it total} non-dimensional sediment concentration. The evolution equations are now
\begin{eqnarray}
\label{eq:non_dim2}
\frac{1}{\Pr}\left(\frac{\partial \bu}{\partial t}+\bu\cdot\nabla\bu\right)=-\nabla p +(T-C)\hat e_z+\nabla^2\bu \, ,\\
\frac{\partial T}{\partial t}+\bu\cdot\nabla T+w=\nabla^2T \, ,\\
\frac{\partial C}{\partial t}+\bu\cdot\nabla C - V \frac{\p C}{\p z}=\tau\nabla^2C \, ,\\
\nabla \cdot \bu=0 \, .
\end{eqnarray}
We consider an initial sediment concentration profile that is invariant in the horizontal directions, and Gaussian in the vertical direction
\begin{equation}
C(x,y,z,0) = A_0 \exp\left[ - \frac{z^2}{ 2 \sigma^2} \right] \, ,
\end{equation}
with initial amplitude $A_0$ and width $\sigma$. We have chosen to use a Gaussian profile because it is mathematically simple and strongly localized while still being very smooth. It may also be a plausible model for the early-time vertical sediment profile in a sediment-laden turbulent gravity current, once it has found its level of neutral buoyancy, but just before the sediments begin settling down.

Although there is no background density ratio parameter $R_0$ in this case, we can still define a $z$- and $t$-dependent density ratio as the ratio of the local non-dimensional temperature gradient to the local non-dimensional sediment concentration gradient:
\begin{equation}
\label{eq: gauss R0}
R(z,t) = \frac{1 + d\bar T/dz }{d\bar C/dz} \, ,
\end{equation}
where, as before, the bars denote a horizontal average.  For convenience, we also define the input parameter $R_{\rm init}$ to be the minimum of $R(z,0)$ over all $z$, which occurs at the lower inflection point of the Gaussian, i.e.
\begin{equation}
R_{\rm init} \equiv \min[R(z,0)]=\frac{\sigma e^{1/2}}{A_0} \, ,
\end{equation}
since the temperature gradient is simply equal to one at time $t = 0$. We therefore see that it is possible to initialize different simulations with the same $R_{\rm init}$ provided $A_0/\sigma$ remains constant. On the other hand, as $A_0$ and $\sigma$ both increase at fixed $R_{\rm init}$, the total amount of sediments present also increases. 

Since $R_{\rm init}$ is the initial local density ratio at the most unstable position in the domain, we expect the fingering instability to develop there first, and to be similar to one that would develop in a constant background gradient setup (see previous sections) with $R_0 = R_{\rm init}$, as long as the initial width of the Gaussian $\sigma$ is much larger than the vertical extent of the fingers. The induced turbulence then disperses the particles, causing the sediment profile to broaden, and to become progressively more stable. At the same time, $R(z,t)$ increases with time everywhere in the domain. Hence, in order to have enough time to observe fully-developed fingering convection over a significant portion of the domain, in what follows we choose a setup where $R_{\rm init}$ is quite small, namely $R_{\rm init} = 1.1$, and where $\sigma$ is reasonably large ($\sigma = 100$). This implies $A_0 \simeq 150$. 


\begin{figure}
\centering
\includegraphics[width=.8\textwidth]{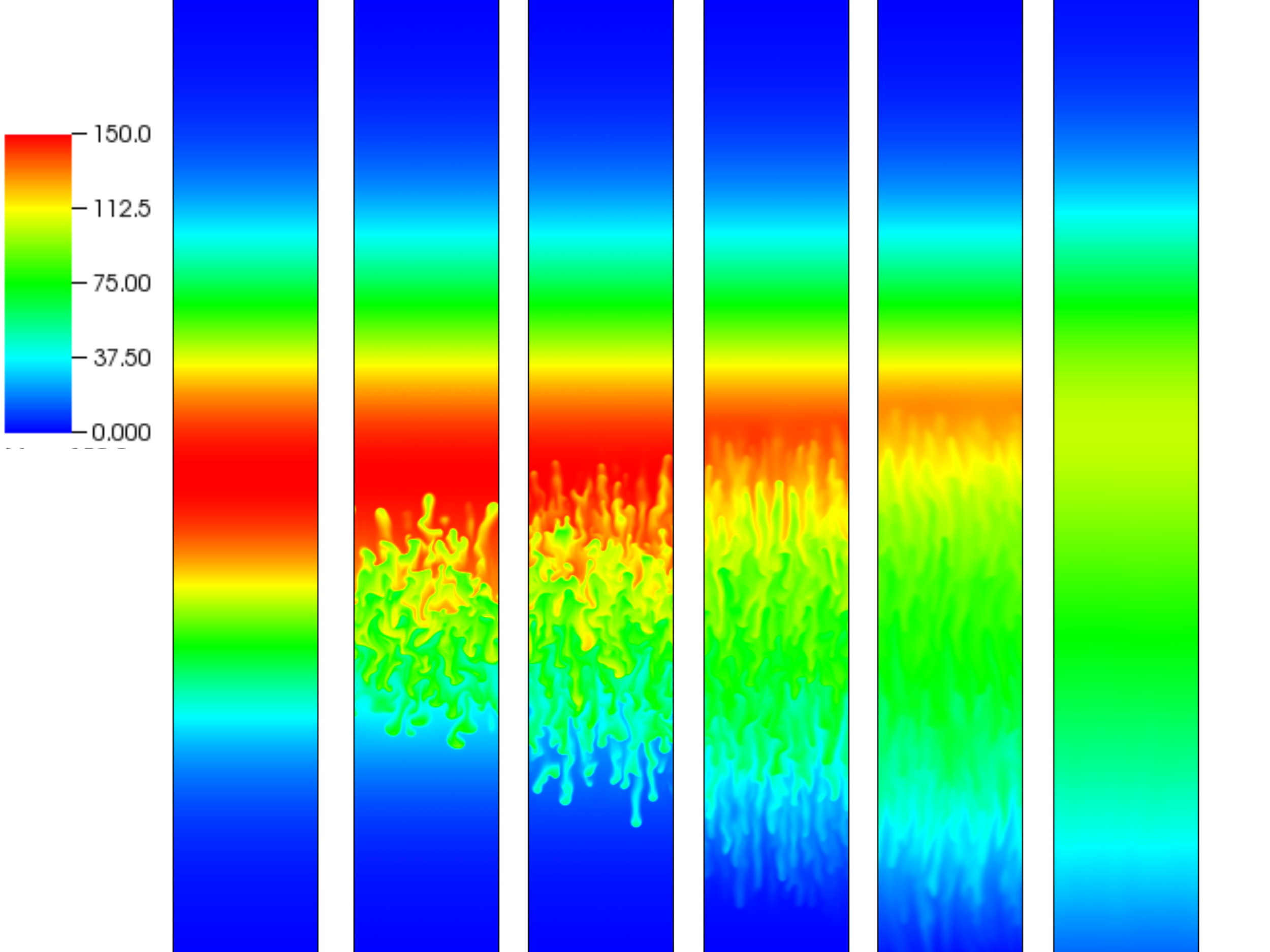}
\caption{Snapshots of the salt concentration in a simulation with $V = 0$, $\Pran = 7$ and $\tau = 0.3$, starting from an initial Gaussian profile with $A_0 = 150$ and $\sigma = 100$. Time increases from left to right, from $t = 0$, $t = 42$, $t = 50$, $t = 104$, $t = 300$ and $t = 5000$.\label{fig:Gaussian}}
\end{figure}

\begin{figure}
\centering
\includegraphics[width=.8\textwidth]{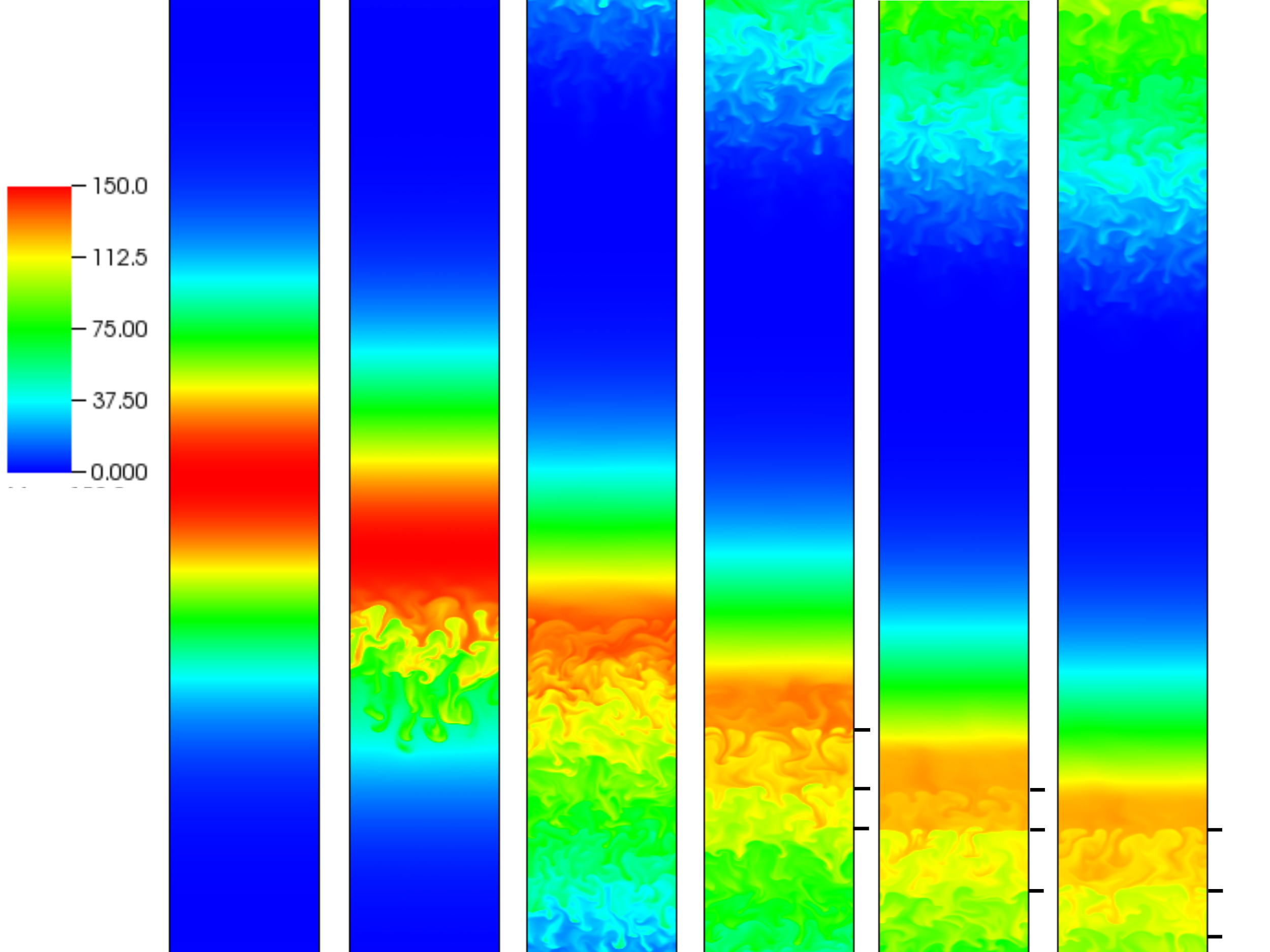}
\caption{Snapshots of the sediment concentration in a simulation with $V = 2.5$, $\Pran = 7$ and $\tau = 0.3$, starting from an initial Gaussian profile with $A_0 = 150$ and $\sigma = 100$. Time increases from left to right, from $t = 0$, $t = 24$, $t = 61$, $t = 91$, $t = 114$ and $t = 128$. The tickmarks roughyl indicate the position of the interfaces when present. \label{fig:stairs}}
\end{figure}

Since our numerical algorithm uses triply-periodic boundary conditions, when picking the height $L_z$ of the 3D domain for each simulation we must ensure that the initial sediment concentration is very close to zero at the top and bottom. Hence we select $L_z = 800$ to ensure that this is the case. To compensate the need for such a large domain in the $z-$direction, we significantly shrink $L_y$ while ensuring that the domain is still wide enough to contain at least 2 wavelengths of the fastest-growing fingering mode \citep{GaraudBrummell2015}.  We therefore take $L_y = 15$. Finally, $L_x$ is set to be $L_x = 100$. All simulations use $\Pran=7$ and $\tau=0.3$. 

To illustrate the qualitative behavior of the results, we compare three simulations with the same initial conditions but different particle settling velocities: $V = 0$, $V = 1$ and $V = 2.5$. 

In the case of non-sedimentary fingering convection ($V = 0$), $C$ can be viewed as the local salinity for instance. Snapshots of the salt concentration at various times (see Figure \ref{fig:Gaussian}) show that the fingering instability first triggers strong mixing near the lower inflection point of the Gaussian, as expected. This reduces the unstable salt gradient, and gradually stabilizes the system. When $R(z,t)$ becomes larger than $1/\tau$ everywhere in the domain, the salt profile becomes stable, and molecular diffusion then takes over. It is interesting to note that layers are not observed to form in this simulation, even though $R_{\rm init}$ is well-into the $\gamma-$unstable regime. This is presumably because the growth rate of meaningful layering modes (i.e. layering modes whose vertical wavelengths are not smaller than the typical finger size) is longer than the time it takes for $R(z,t)$ to exceed the critical threshold $R_{\rm crit} \simeq 1.5$ beyond which the $\gamma-$instability disappears. 

\begin{figure}
\centering
\includegraphics[width=\textwidth]{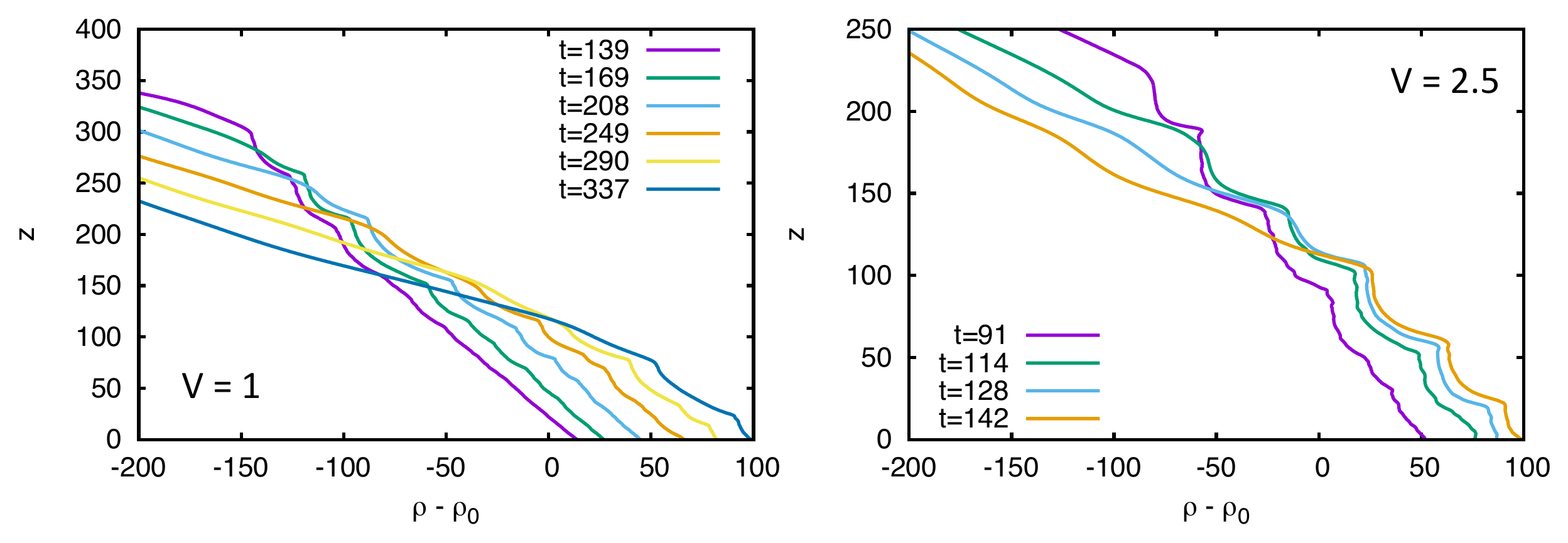}
\caption{Mean density profiles as a function of time and $z$ for simulations with $\Pran = 7$ and $\tau = 0.3$, starting from an initial Gaussian profile with $A_0 = 150$ and $\sigma = 100$. On the left, $V =1$ and on the right, $V = 2.5$. Only the lower portion of the computational domain is shown. Although layering modes are clearly growing in the $V=1$ case, they are not strong enough to cause convective overturning (characterized by regions where $\bar \rho$ increases upward). Convective layers do appear, on the other hand, in the $V = 2.5$ case. \label{fig:densitystairs}}
\end{figure}

In the case of sedimentary fingering convection with relatively rapid sedimentation ($V=2.5$), we may expect a staircase to form through the $\gamma-$instability since sedimentation causes $R_{\rm crit}$ to be greatly increased (see Section \ref{sec: basic flux}). We find that this is indeed the case: snapshots of the sediment concentration at various times (see Figure \ref{fig:stairs}) now show the development of layers. To see the layers more clearly, we plot the horizontally-averaged density profile in the lower half of the computational domain in Figure \ref{fig:densitystairs}. This confirms that what appears to be layers in the simulation snapshots are indeed part of a well-defined density staircase, with convectively unstable layers separated by stable interfaces. As in the more idealized simulations presented in Section \ref{sec: Comparing results to linear sim}, we see that the interfaces migrate slowly upward. By contrast, however, no mergers are seen, but this is perhaps because the merger timescale may be longer than the time it takes for the peak of the sediment profile to settle down through the staircase. Instead, we see new layers appear and strengthen near the bottom of the staircase, as they disappear from the top. 

Finally, with an intermediate settling velocity (here, $V = 1$), we may expect a staircase to form through the new sedimentary layering instability discussed in Section \ref{sec: sedlayer}. In this case, however, we cannot see layers appear in simulation snapshots (not shown) as clearly as they do in the case with $V = 2.5$. Instead, the system looks qualitatively similar to the third panel of Figure \ref{fig:stairs} at all times. Inspection of the density profiles on the other hand (see Figure \ref{fig:densitystairs}) clearly show that there {\it is} a layering instability, but the latter does not seem to grow to sufficiently large amplitudes to cause convective overturning. We speculate that this is because the layer growth rate in the $V = 1$ simulations is smaller than in the $V =  2.5$ simulation, and the layers do not have time to fully form before the peak of the sediment profile has settled through the growing layering modes.

\section{Summary and prospects} \label{sec: Discussion}

In this investigation, we have found that sedimentary fingering convection is more likely to undergo layer formation than normal fingering convection, for two reasons. First, rapid settling ($V>1$) extends the region of parameter space that is susceptible to the $\gamma-$instability \citep{radko2003mechanism}, by significantly affecting the turbulent temperature and particle fluxes. Furthermore, a new instability appears that exists even when the $\gamma-$instability does not operate. A sufficient condition for the new sedimentary layering instability is that $R{\rm Nu}$ be a decreasing function of the density ratio, which seems to be nearly always the case for reasonable parameter values. Its predicted growth rate scales roughly with the square of the settling velocity $V$, while the wavelength of the fastest-growing layering mode scales with $1/V$.  Hence, if $V$ is too small, the layering modes cannot fit in the domain and/or grow too slowly to be relevant -- but for large enough $V$, the instability could in principle be significant. Direct numerical simulations support our findings qualitatively, but also reveal that the growth rates of both layering instabilities are poorly predicted by theory at large density ratios. This could be attributed to the strong vertical coherence of the fingers at these parameters, while the mean field theory developed assumes that the fluxes are inherently local. Future work on the topic should attempt to improve  the theory to match the observed layering mode growth rates and, when applicable, to provide the initial height of the forming layers.
 
Given the aforementioned discrepancies, and given the fact that most of our numerical experiments have been run with a diffusivity ratio of $\tau = 1/3$, which is much larger than the one relevant for natural sedimentary fingering flows, it is premature to make observational predictions about layer formation in real geophysical conditions (i.e. river outflows, volcanic plumes or marine snow). To do so in any reliable fashion, one would need (1) a theory to reliably predict the turbulent fluxes for much lower values of $\tau$, for $V \neq 0$, which could for instance be obtained from an extension of the work of \citet{RadkoSmith2012}, and (2) to improve on the current mean field theory presented in this paper. Rather, our goal here  was to present the theory and preliminary numerical simulations. 

Nevertheless, we venture here as far as providing a naive guess of the dimensional layer heights and growth rates in sedimentary fingering convection using the available data, focussing on the cases that are $\gamma-$stable in the limit where $V \rightarrow 0$. 

Let us first discuss the case of the sedimentary layering instability. Assuming that (1) the turbulent fluxes at $\Pr = 7$, $\tau = 0.01$ obtained by \citet{Traxler2011a} are decent estimates for the turbulent fluxes for sedimentary fingering convection with $\tau \ll 1$, $V \ll 1$, and moderate values of the density ratio $R_0$, and (2) using the asymptotic formulas given in (\ref{eq: ass exp}) with these fluxes, then we find typical non-dimensional layer growth rates in the $\gamma$-stable region of parameter space, for $4<R_0<10$, ranging from $10^{-5}V^2$ to $10^{-4}V^2$, and vertical wavenumbers of the order of $10^{-2}V$. Using a fiducial value for $V$ of $V \simeq 0.1$, a unit length $[l]  = d \simeq 0.01$m, and a unit time $[t] = d^2/\kappa_T \simeq 10^3$s (see Section \ref{sec: The Model}), we find growth timescales of the order of $10^{10}$s which corresponds to a few hundred years, and layer heights of a few tens of meters. Taken at face value, the estimated growth timescale is likely too long for this instability to be relevant in oceans and lakes. Even accounting for the yet unexplained discrepancies between theory and DNS, which show that the theory under-predicts the growth rates by factors up to 10, we would still only achieve growth rates of the order of 10 years. It remains to be determined, however, whether this instability could be relevant for suspended particles in air, or for sedimentary fingering convection in more exotic scenarios such as in planetary atmospheres or stellar interiors for instance.  

By contrast, the extension of the $\gamma-$instability range for sedimentary fingering convection with rapid settling ($V \ge 2$) is a much more promising avenue towards layer formation. Again, we cannot give strict predictions for the growth rates of $\gamma-$unstable layering modes for sediments in water, since we do not yet have turbulent flux measurements at low diffusivity ratios and non-dimensional settling velocities of order one or higher. However, if we naively assume, for the purpose of this exercise, that we can directly apply the results of our DNS at $R_0 = 2$, $V=2$, $\Pr = 7$ and $\tau = 1/3$, then we find that layers should grow at a non-dimensional rate of about $10^{-2}$ to $10^{-1}$, with non-dimensional initial heights about $50$ to $100$ (see Table 2), which corresponds to dimensional growth timescales of about a day to  a week, and initial heights of about half a meter to a meter. How much this naive estimate would need to be revised at lower values of $\tau$ remains to be determined, but if it does not change too much, then indeed layer formation in sedimentary fingering convection in water, for $V \ge 2$, could be quite ubiquitous thanks to the $\gamma$-instability. 
 
Beyond the layer formation process, much still remains to be done in terms of modeling the structure of sediment staircases. For instance, while convection in thermohaline staircases is driven by the differential transport of heat and salt across the interfaces, sedimentation can also play an important role in the process for $V \ge 1$. Quantifying this effect will be the subject of future work, and will hopefully shed light on the reason why interfaces move upward, rather than downward. Another important avenue for future research will be to further investigate the overall effect of layer formation on the particle sedimentation rates. As we have seen throughout this paper, layer formation has a tendency to enhance the downward turbulent fluxes of both temperature and sediment over that of the basic fingering fluxes \citep[see also][]{radko2003mechanism,Stellmach2011}, sometimes significantly when layers merge with one another and grow in height. These mergers were only seen in the idealized simulations that use constant background temperature and concentration gradients, however. In the somewhat more realistic situations explored in Section \ref{sec: Gaussian}, where we followed the evolution of a narrow horizontal sediment layer, the enhancement of downward transport was relatively weak because layers did not merge after they formed. Whether this will always be the case remains to be determined, but the answer will clearly have a significant impact on the vertical transport of sediments in stratified fluids. 

\acknowledgements

The authors thank S. Stellmach for granting us the use of his code, and M. Wells for interesting discussions. This work is supported by NSF CBET-1437275 for JR and PG,  NSF CBET-1438052 to EM, as well as by grant DN NSWC N00174-16-C-0013 to EM.  Financial support was also provided by the Saudi Arabian Oil Company (Saudi Aramco) to AA. This work used the Hyades supercomputer, funded by an NSF MRI grant, and the Extreme Science and Engineering Discovery Environment (XSEDE), which is supported by National Science Foundation grant number ACI-1053575.

\bibliography{references_layer_draft}
\bibliographystyle{jfm}

\end{document}